%% file: RelTimeDil_final.tex
\documentclass[accepted=2020-08-07,aps,showpacs,superscriptaddress,twocolumn,nofootinbib,a4paper]{quantumarticle}
\pdfoutput=1
\usepackage{graphicx}  % needed for figures
\usepackage{amsfonts}
\usepackage{dsfont}
\usepackage[breaklinks=true,colorlinks=true,linkcolor=blue,urlcolor=blue,citecolor=blue]{hyperref}
\usepackage{comment}
\usepackage{color}
\usepackage[makeroom]{cancel}
\usepackage{footnote}

\renewcommand{\[}{\begin{equation}}
\renewcommand{\]}{\end{equation}}

\newcommand{\tr}{\mathrm{tr}}

\newcommand{\dom}{\textup{Dom}}

\newcommand{\ArXiv}[1]{}%{#1} % replace with {#1} %{} 

\input{MischaCommands}

\begin{document}

\title{Universal quantum modifications to general relativistic time dilation in delocalised clocks}

\author{Shishir Khandelwal}
\affiliation{Institute for Theoretical Physics, ETH Z\"urich, Switzerland}
\affiliation{Group of Applied Physics, University of Geneva, Switzerland}
\author{Maximilian P. E. Lock}
\affiliation{Institute for Quantum Optics and Quantum Information (IQOQI), Vienna, Austria}
\author{Mischa P. Woods}
\affiliation{Institute for Theoretical Physics, ETH Z\"urich, Switzerland}
\affiliation{Department of Computer Science, University College London, United Kingdom}

%\date{\today}

\begin{abstract}
The theory of relativity associates a proper time with each moving object via its world line. In quantum theory however, such well-defined trajectories are forbidden. After introducing a general characterisation of quantum clocks, we demonstrate that, in the weak-field, low-velocity limit, all ``good'' quantum clocks experience time dilation as dictated by general relativity when their state of motion is classical (i.e. Gaussian). For nonclassical states of motion, on the other hand, we find that quantum interference effects may give rise to a significant discrepancy between the proper time and the time measured by the clock. The universality of this discrepancy implies that it is not simply a systematic error, but rather a quantum modification to the proper time itself. We also show how the clock's delocalisation leads to a larger uncertainty in the time it measures --- a consequence of the unavoidable entanglement between the clock time and its center-of-mass degrees of freedom. We demonstrate how this lost precision can be recovered by performing a measurement of the clock's state of motion alongside its time reading.
\end{abstract}

\maketitle

%%%%%%%%%%%%%%%%%%%%%%%%%%%%%%%%%%%%%%%%%%%%%%%%%%%%%%%%%%%%%%%%%

\section{Introduction}
One of the most important programs in theoretical physics is the pursuit of a successful theory unifying quantum mechanics and general relativity. Arguably, many of the difficulties arising in this pursuit stem from a lack of understanding of the nature of time~\cite{isham1993canonical,kuch1993,anderson2017quantum}, particularly the conflict between how it is conceived in the two theories~\cite{lock2017relativistic}. For example, in general relativity a given object's proper time is defined geometrically according to that object's world line, but according to quantum mechanics such well-defined trajectories are impossible. How do we then assign a proper time to the delocalised objects described by quantum theory?

Confusion over the nature of time is not limited to the context of quantum gravity. Indeed, even in non-relativistic quantum mechanics, there is much to clarify. In the latter, time is not treated on the same footing as other observable quantities in the theory --- it is a parameter, rather than being represented by a self-adjoint operator. This aspect of the theory of quantum mechanics troubled its founders; Wolfgang Pauli, for example, famously noted the impossibility of a self-adjoint operator corresponding to time~\cite{pauli1958allgemeinen}. Specifically, a quantum observable with outcome $t\in\rr$ equal to the time parameterising the system's evolution, can only be achieved in the limiting case of infinite energy, and is therefore unphysical, as we discuss in Sec.~\ref{sQuantClocks}. It is however possible to construct self-adjoint operators whose outcome is approximately $t$, with an error that can in principle be made arbitrarily small, in systems of finite energy (see e.g.~\cite{woods2018autonomous}).

If one takes an operationalist viewpoint, one can draw conclusions about time by discussing the behaviour of clocks. This underlies Einstein's development of special relativity~\cite{einstein1905elektrodynamik}, the prototypical example of operationalism~\cite{bridgman1927logic}. In a quantum setting, the clocks must of course be quantised, leading naturally to a time operator. This operational approach has revealed fundamental limitations to time-keeping in non-relativistic quantum systems~\cite{buzek1999optimal,woods2018autonomous,erker2014quantum,woods2018quantum}, and novel effects in a variety of relativistic settings~\cite{salecker1958quantum,lindkvist2014twin,lorek2015ideal,lock2019quantum,anastopoulos2018equivalence}.

\begin{figure}[h]
	\centering
	\includegraphics[width=0.99\columnwidth]{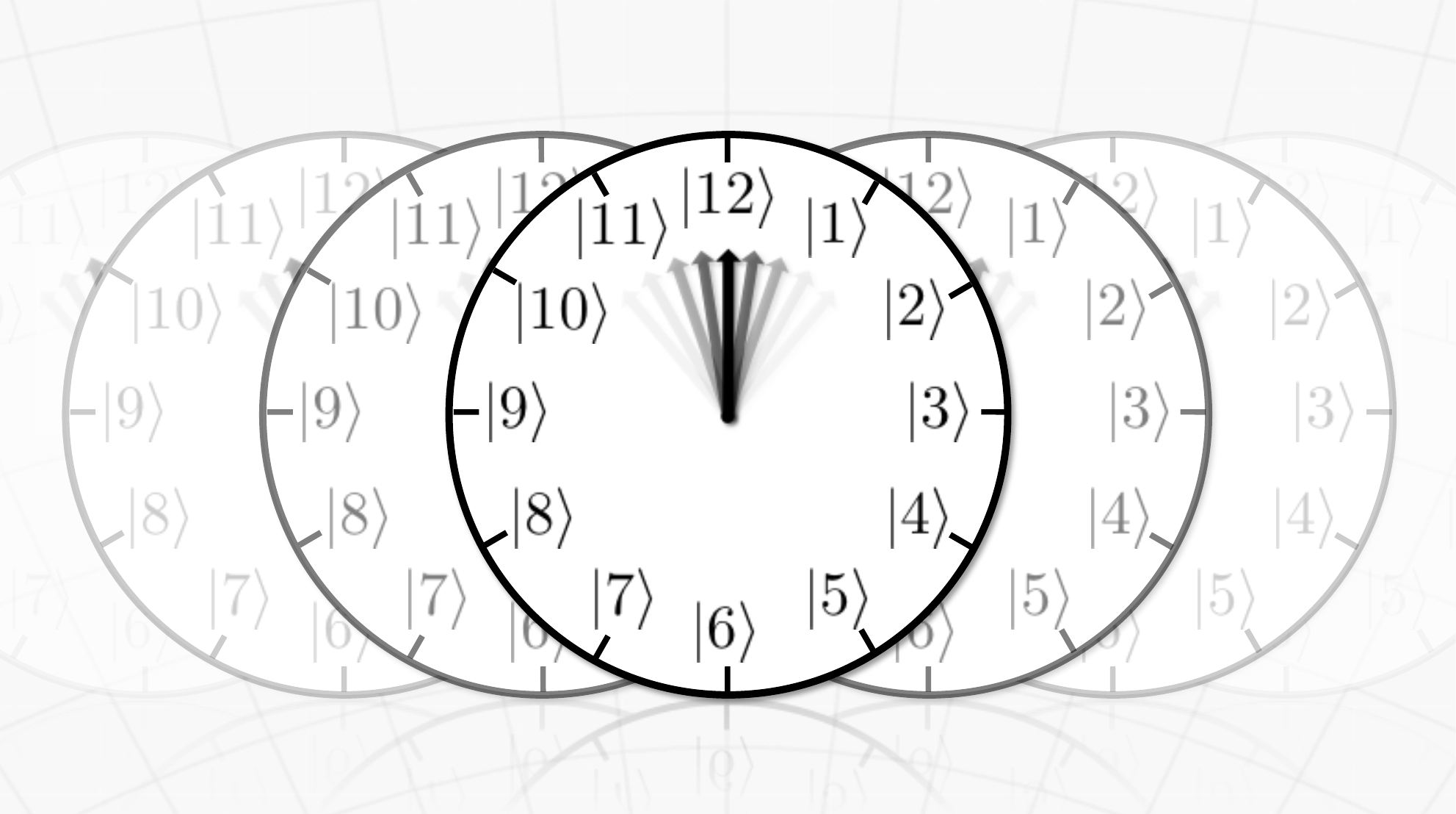}
	\caption{Depiction of a spatially-delocalised quantum clock exibiting temporal indeterminacy. In the theory of relativity, the time measured by a clock is defined geometrically by integrating along a well-defined spacetime path, and the clock may be arbitrarily precise and accurate. This contrasts with time in quantum mechanics, where a fully quantum treatment requires clocks to be delocalised in space and experience an indeterminacy in their time. They cannot be meaningfully assigned a single trajectory. Relativity dynamically couples the temporal and kinematic quantum degrees of freedom, resulting in time measurements which are influenced by the delocalised spatial trajectory.} \label{fDelocalisedSpacetimeCartoon}
\end{figure}

The present work concerns slowly-moving generic quantum clocks embedded in a weakly-curved spacetime. We are interested in the phenomenon of time dilation, and how quantizing the clock may result in predictions distinct from those of general relativity alone. We take the operational viewpoint noted above, associating time with the outcomes of measurements performed on generic quantum clocks. In this setting, in contrast with the study of clocks in non-relativistic quantum mechanics, we must consider the position and momentum of the clock. Quantum theory dictates that it be subject to some degree of spatial delocalisation in addition to its ``temporal delocalisation'' (i.e. the indeterminacy of its time-reading), as depicted in Fig.~\ref{fDelocalisedSpacetimeCartoon}.

Relativistic effects manifest in this quantum setting via a coupling between the clock's kinematic and internal (in our case, time-measuring) degrees of freedom. The form of this coupling has been derived in a number of different ways: from the relativistic dispersion relation and mass-energy equivalence~\cite{zych2017quantum}, from the Klein-Gordon equation in curved spacetime~\cite{lammerzahl1995hamilton} (or its corresponding Lagrangian density~\cite{anastopoulos2018equivalence}), and as a consequence of classical time dilation between the clock and the laboratory frame of reference~\cite{zych2011quantum}. It has been used to predict a relativistically-induced decoherence effect~\cite{pikovski2015universal}, and a corresponding reduction in the visibility of quantum interference experiments~\cite{zych2011quantum}. This approach treats spacetime classically, making no reference to a quantum theory of gravity. It is applicable only in the low-energy, weak-gravity limit, so that we may consider relativistic effects as perturbative corrections to non-relativistic quantum mechanics.

We begin by introducing a way of characterising the accuracy of a generic quantum clock, before describing how to incorporate the phenomenon of time dilation into quantum dynamics. This is followed by a description of the average time dilation experienced by a quantum clock for two different states of motion. We then discuss how the incorporation of relativity increases the uncertainty associated with the measurement of time, and how this can be mitigated by gaining knowledge of the clock's location/motion, illustrating this with a numerical example. We finish with a discussion of our results and the questions that arise from them.

%%%%%%%%%%%%%%%%%%%%%%%%%%%%%%%%%%%%%%%%%%%%%%%%%%%%%%%%%%%%%%%%%

\section{Non-Relativistic Quantum Clocks}\label{sQuantClocks} 

Before we discuss relativistic effects, it is convenient to introduce a characterisation of the extent to which a given quantum clock measures time accurately. We associate the relevant degree of freedom with a Hilbert space $\mathcal{H}_{\cl}$, and consider an initial clock state ${\rho}_{\cl}(0)$ which evolves according to the Schr\"odinger equation, with Hamiltonian $\hat H_\cl$. We use the term \emph{\exttime{} time} to refer to the parameter appearing in the Schr\"odinger equation. At \exttime{} time  $t$, a measurement is performed, which corresponds to a Positive-Operator Valued Measure (POVM) $\{\hat{F}(s)\}_{s\in\,\mathcal{S}}$ with $\mathcal{S} \subseteq \rr$. We refer to the outcome $s$ of this measurement as the \emph{clock time}. A quantum clock is then defined by the tuple $\big\lbrace \{\hat F(s)\}_{s \in\,\mathcal{S}}, \hat{H}_{\cl}, {\rho}_{\cl}(0) \big\rbrace$. In this article however, it suffices to consider $\big\lbrace \hat{T}_{\cl}, \hat{H}_{\cl}, {\rho}_{\cl}(0) \big\rbrace$, where ${ \hat T_\cl := \int_{\mathcal{S}} s \, \hat F(s ) ds }$ is the first-moment operator of the POVM, which determines the expected value of the measurement. In the case where the POVM is a projection-valued measure (in other words, where the clock times are perfectly distinguishable) the definition of $\hat T_\cl$ above is simply the spectral decomposition of the self-adjoint operator corresponding to the measurement. For simplicity, we choose the \exttime{} time $t$ to be such that $\langle \hat T_{\cl} \rangle (0) = 0$.

If a quantum clock satisfies the Heisenberg form of the canonical commutation relation, $\big[ \hat{T}_{\cl}, \hat{H}_{\cl} \big]=i\,\id_\cl$,\footnote{Here and throughout this \doc, we use units such that $\hbar=1$. The symbol $\id_\cl$ denotes the identity operator on $\mathcal{H}_{\cl}$.} the mean clock time will exactly follow the \exttime{} time, i.e. $\langle \hat T_{\cl} \rangle(t) = t \text{ mod } T_0$ $\forall t$, for some (possibly infinite) clock period  $T_0>0$. Furthermore, the standard deviation of such clocks is constant and determined by the initial state which can be arbitrarily small. We refer to such a clock as \emph{idealised}. If one demands only this commutation relation, one can construct an idealised clock on a dense subset of the Hilbert space, with finite period $T_0$ and with $\hat{H}_{\cl}$ bounded below (e.g. the phase operators in~\cite{garrison1970canonically}). If we further impose canonical commutation relations of the Weyl form, and consider measurements corresponding to projection-valued measures, we necessarily have that $\hat{H}_{\cl}$ is unbounded below (Pauli's theorem~\cite{pauli1958allgemeinen,srinivas1981time}), which is clearly unphysical, and ${T_0=\infty}$. These are both examples of \emph{covariant} POVMs~\cite{Holevo,buzek1999optimal}, i.e. those satisfying the covariance property $\hat F(s)=\me^{-\mi s\hat H_\cl} \hat F(0)\me^{\mi s\hat H_\cl}$, $s\in\rr$; where in the case that $\mathcal{S}$ is bounded, $\hat F(s)$ has been periodically extended to $\rr$. Furthermore, they satisfy the canonical commutator relation of the Weyl form when $\mathcal{S}$ is unbounded, and of the Heisenberg form if ${\rho}_{\cl}(0)$ and $\hat{F}(0)$ have disjoint support (see \App~\ref{app:POVMs} for proof).

To derive results that apply universally to all quantum clocks, regardless of whether they exhibit the idealised behaviour discussed above, we first quantify the extent to which a given clock deviates from an idealised one. To this end, we introduce a clock's \emph{error operator} $\hat{E}(t)$; for any clock characterised by $\big\lbrace \hat{T}_{\cl}, \hat{H}_{\cl}, {\rho}_{\cl}(0) \big\rbrace$, we define $\hat{E}(t)$ via 
\begin{equation} \label{eEdefn}
- i \left[\hat{T}_{\cl} , \hat{H}_{\cl} \right] {\rho}_{\cl}(t) = {\rho}_{\cl}(t) + \hat{E}(t)
\end{equation}
where ${\rho}_{\cl}(t)=e^{- i t \hat H_\cl} {\rho}_{\cl}(0)\, e^{i t \hat H_\cl}$. %\footnote{In the case of projective measurements \mpw{and/or covariant POVMs,} $\hat E(t)$ also characterises the errors in said higher moments. See \App~\ref{app:POVMs}.}
While there are a variety of ways one might characterise the deviations of any clock, it will turn out that for our purposes, the definition of $\hat E(t)$ is particularly convenient. As initial motivation, observe that the error operator $\hat{E}(t)$ allows us to quantify the discrepancy between the average clock time and the \exttime{} time $t$ via
\begin{equation} \label{eFreeMeanTime}
\langle \hat{T}_{\cl} \rangle_\NR(t) = t + \int_{0}^{t} dt' \tr \left[ \hat{E}(t') \right]
\end{equation}
where the subscript NR denotes that the clock does not account for relativistic effects (i.e. it evolves under the action of $\hat H_{\cl}$ alone), and we have assumed for convenience that $\langle \hat{T}_{\cl} \rangle_\NR(0)=0$. %An identical equation holds in the case of a more general measurement, namely one represented by a Positive-Operator Valued Measure (POVM), as detailed in \App~\ref{app:POVMs}. 
For an idealised clock, we have $- i\big[ \hat{T}_{\cl}, \hat{H}_{\cl} \big]=\id_\cl$, and therefore $\hat{E}(t)=0$~$\forall t$. We therefore call a clock \emph{good} when $\hat E(t)$ is small in norm relative to the other quantities involved.  

An interesting case is when the clock is a $d$-dimensional spin system, with evenly-spaced energy eigenvalues (corresponding to the spin-projection states). Constructing $\hat{T}_{\cl}$ such that its eigenvectors form an appropriate mutually-unbiased basis with respect to the spin-projection states, and choosing ${\rho}_{\cl}(0)$ to be one of the eigenvectors of $\hat{T}_{\cl}$, one arrives at the well-known Salecker-Wigner-Peres clock~\cite{salecker1958quantum,peres1980measurement} --- a quantum version of the common dial clock (see Fig.~\ref{fDelocalisedSpacetimeCartoon}). In this case, $\hat E(t)$ is not always small; one has $ \mathrm{tr} \big[ \hat{E}(t) \big] =-1$ at regular intervals of \exttime{} time, regardless of the dimension $d$ or energy. This arises due to the clock states' lack of coherence in the eigenbasis of $\hat{T}_{\cl}$; specifically $[\hat T_{\cl},{\rho}_{\cl}(t)]$ is zero at regular intervals of \exttime{} time. This issue can be removed by choosing ${\rho}_{c}(0)$ to have some quantum coherence in this basis, as in the \emph{Quasi-Ideal} clock~\cite{woods2018autonomous}. In the latter case, $\| \hat E(t)\|_{1}$ is exponentially small in both the dimensionality $d$ and the mean energy of the clock $\forall t$. This is a consequence of the Quasi-Ideal clock's approximately %non-dispersiveness 
dispersionless dynamics in the eigenbasis of $\hat{T}_{\cl}$. %\cc{Moreover, the uncertainty of the clock time is constant in this case, up to a similar exponentially small correction~\cite{woods2018autonomous}.}\mpw{Mention t-incoherent states and ref. your paper with Alv} 
In addition to the Quasi-Ideal clock, the Salecker-Wigner-Peres and  qubit-phase clocks, which do not posses this property, are discussed in \App~\ref{app:egs}.

\section{Relativistic Time Dilation} 

Classically, in the post-Newtonian approximation of general relativity, the proper time $\tau$ experienced by an observer is determined by~\cite{weinberg1972gravitation}
\begin{equation} \label{ePropTimeDifferential}
d \tau = d\tilde{t} \left( 1 - \frac{v^{2}}{2 c^{2}} + \frac{\Phi(r)}{c^{2}} \right),
\end{equation} 
to first order in $v^{2}/c^{2}$ and $\Phi/c^{2}$, where $v$ is the observer's velocity, $r$ is the distance from a gravitating body of mass~$M$, $\Phi(r)=-G M/r$ is the Newtonian gravitational potential, and $\tilde{t}$ is the proper time of a fictional observer at rest at $r=\infty$. We take the common approximation of the Newtonian potential around a point $r_{0}$, i.e. $\Phi(r) \approx \Phi(r_{0})+g x$, where $g$ is the gravitational acceleration at the point $r=r_{0}$ (e.g. $g=9.81$\!~m\!~s${}^{-2}$ at the surface of the Earth), and $x:=r-r_{0}$. The reference frame corresponding to the laboratory is then given by $(t,x)$, where the laboratory time $t$ is the proper time of an observer at rest at $r=r_{0}$. Then, for a classical clock with initial position $x_{0}$ and velocity $v_{0}$, Eq.~(\ref{ePropTimeDifferential}) allows us to relate the clock's proper time $\tau$ to the \exttime{} time:
\begin{equation} \label{ePropTime}
\tau = \left[ 1 - \frac{v_{0}^{2}}{2 c^{2}} + \frac{g x_{0}}{c^{2}} + \frac{v_{0}gt}{c^{2}} - \frac{1}{3} \left( \frac{g t }{c} \right)^{2} \right] t .
\end{equation}

If we now consider the clock to be subject to the laws of quantum mechanics, we must describe how its temporal degree of freedom interacts with its kinematic degrees of freedom. To that end, we consider a total Hilbert space $\mathcal{H}=\mathcal{H}_{\cl} \otimes \mathcal{H}_{\ki}$, where the space $\mathcal{H}_{\ki}$ corresponds to the kinematic variables in one dimension, namely position $\hat{x}$ and momentum $\hat{p}$, with $[\hat{x},\hat{p}]=i\,\id_{\ki} $. The coupling between the temporal and kinematic spaces (which can be understood as a consequence of mass-energy equivalence~\cite{zych2018quantum}) is then given by the interaction Hamiltonian~\cite{lammerzahl1995hamilton,pikovski2015universal,zych2017quantum}
\begin{equation} \label{eIntHam}
\hat{H}_{\cl\ki} = \hat{H}_{\cl} \otimes \left(- \frac{\hat{p}^{2}}{2m^{2}c^{2}} + \frac{g \hat{x}}{c^{2}} \right) ,
\end{equation}
where $m$ is the mass of the clock, and where we continue the approximation in Eq.~\eqref{ePropTimeDifferential}, neglecting terms proportional to $1/c^{4}$ (see \App~\ref{app:coupling}). The two terms in Eq.~(\ref{eIntHam}) represent the lowest-order contributions to the time dilation due to motion (c.f. $v^{2}/2c^{2}$ in Eq.~(\ref{ePropTimeDifferential})) and gravity respectively. The clock is then subject to the total Hamiltonian $\hat{H} = \hat{H}_{\cl} + \hat{H}_{\ki} + \hat{H}_{\cl\ki}$, with ${\hat{H}_{\ki} = mc^{2} + m g \hat{x} + \hat{p}^{2}/2m - \hat{p}^{4}/8m^{3}c^{2}}$, where the final term is the usual lowest-order relativistic correction to the kinetic energy.

\section{Time Dilation in Quantum Clocks}

We assume the initial state of the clock to be uncorrelated across $\mathcal{H}_{\cl}$ and $\mathcal{H}_{\ki}$, i.e. that $\rho_{\cl\ki}(0)=\rho_\cl(0)\otimes\rho_\ki(0)$ for some $\rho_\cl(0)$ and $\rho_\ki(0)$. After evolving for an amount of \exttime{} time $t$, the average time dilation experienced by the clock (i.e. the average clock time) is
\begin{equation} \label{eQuantTimeDilation}
\langle \hat{T}_{\cl} \rangle ( t ) =  \langle \hat{T}_{\cl} \rangle_\NR(t) +  t R(t) \left\lbrace 1 + \mathrm{tr} \left[ \hat{E}(t) \right] \right\rbrace ,
\end{equation}
where $R(t):= \tr \left[ \left( - \frac{\hat p^2}{2 m^2 c^{2}} + \frac{g \hat x}{c^{2}}  + \frac{\hat{p}gt}{mc^{2}} - \frac{g^2t^{2}}{3c^{2}} \right) \rho_{\ki}(0) \right]$ is determined by the initial kinematic state. Recall that $\hat{E}(t)$ was defined according to the non-relativistic evolution of the clock (i.e. under $\hat{H}_{\cl}$ alone). %In \App~\ref{app:POVMs}, we show how a similar equation to Eq.~\ref{eQuantTimeDilation} holds for a general POVM.
We now consider two possibilities for $\rho_\ki(0)$ | a classical (i.e. Gaussian) and a nonclassical state. 

\subsection{A Gaussian state of motion} \label{sClassMotion}
We take the initial kinematic state $\rho_\ki(0)$ to be a pure Gaussian wavepacket, denoting the mean momentum by $\bar{p}_{0}$, the standard deviation of the momentum by $\sigma_{p}$, and the mean position by $\bar{x}_{0}$. This is the most classical choice in the sense that such states are the only ones with a non-negative Wigner function~\cite{hudson1974wigner}, and saturate the position-momentum uncertainty relation. We then have
\begin{equation}
R(t)=- \frac{\bar{p}_{0}^{2}+\sigma_{p}^{2}}{2 m^{2} c^{2}} + \frac{g \bar{x}_{0}}{c^{2}} + \frac{\bar{p}_{0}gt}{m c^{2}} - \frac{1}{3} \left( \frac{g t }{c} \right)^{2}.\label{eq:R t Gaussina}
\end{equation}
In this case, setting $\hat{E}(t)=0$ and using Eqs.~(\ref{eFreeMeanTime}) and~(\ref{eQuantTimeDilation}), one finds that the average time measured by an idealised quantum clock is identical to the expectation value of the classical proper time (Eq.~(\ref{ePropTime})) for an observer with a momentum following the same (i.e. Gaussian) probability distribution. A non-idealised clock will exhibit a non-relativistic quantum correction (the second term in Eq.~(\ref{eFreeMeanTime})) as well as a contribution arising from both relativity and quantum mechanics (the final term in Eq.~(\ref{eQuantTimeDilation})). However, since both these effects are proportional to $\mathrm{tr} \big[ \hat{E}(t) \big]$, they can be made arbitrarily small, for example by using the Quasi-Ideal clock discussed in Sec.~\ref{sQuantClocks} with an appropriately high dimensionality and mean energy~\cite{woods2018autonomous}. Interestingly, for the Salecker-Wigner-Peres clock discussed Sec.~\ref{sQuantClocks}, $\mathrm{tr} \big[ \hat{E}(t) \big]=-1$ whenever the clock is in an eigenstate of $\hat{T}_{\cl}$, exactly cancelling the usual classical relativistic effect in Eq. \eqref{eQuantTimeDilation} (see \App~\ref{app:egs}). This property of the Saleker-Wigner-Peres clock states has also been identified as the cause for suboptimal performance in other areas, related to quantum error correction with clocks \cite{WoodsAlv2019}.

\subsection{A superposition of heights} \label{sQuantMotion}
\begin{figure}
  \centering
    \includegraphics[width=0.99\columnwidth]{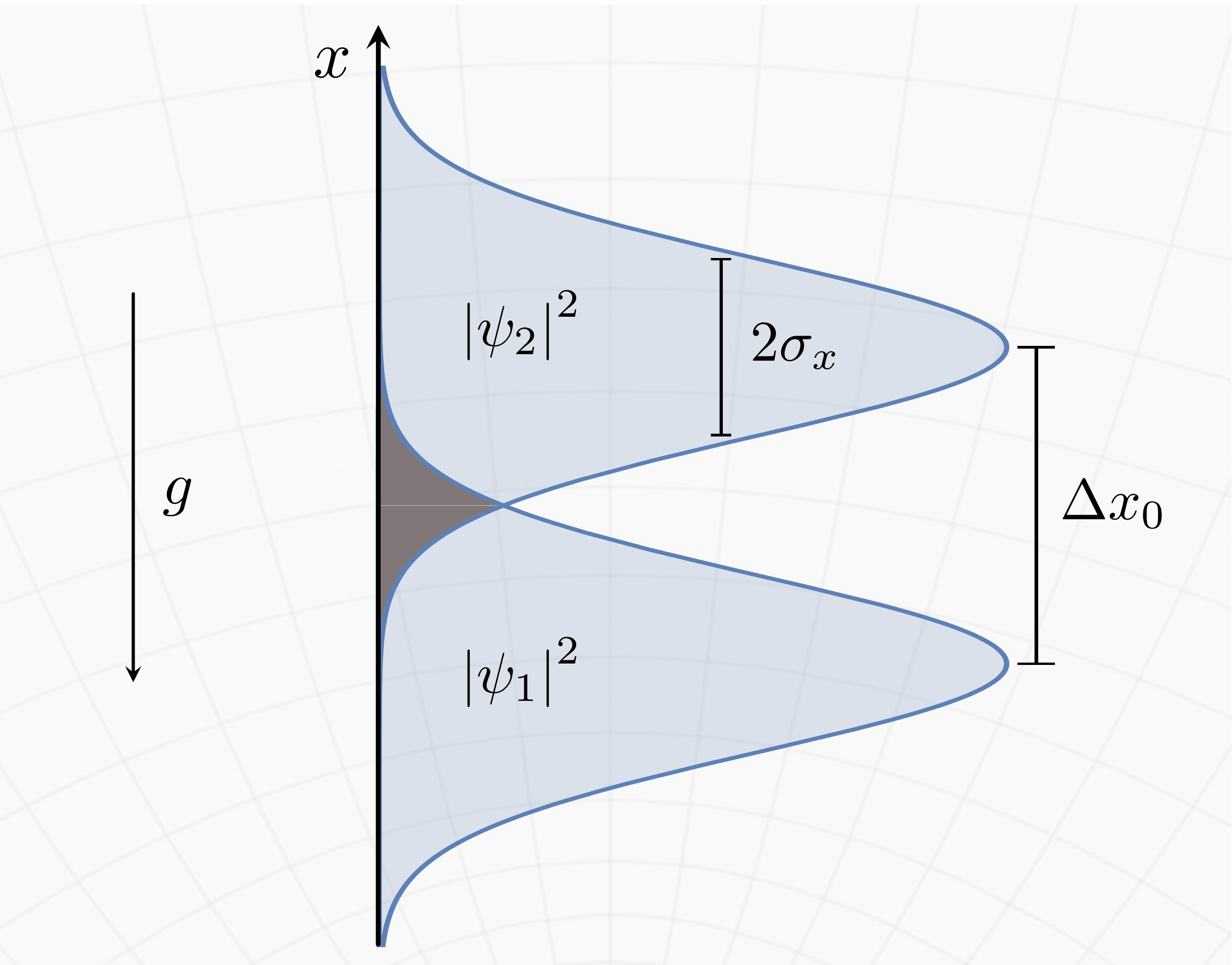}
      \caption{A cartoon of the (magnitude squared of) the wavefunctions used to construct delocalised initial (kinematic) states of the clock.} \label{fSuperposCartoon}
\end{figure}

A consideration of a nonclassical initial kinematic state can result in a significant modification of the classical time-dilation effect found in Sec.~\ref{sClassMotion}. We consider an initial kinematic state constructed by superposing two Gaussian wavepackets with different mean initial heights, namely
\begin{equation}
\ket{\psi}_{\ki} = \frac{1}{\sqrt{N}} \left( \sqrt{\alpha} \ket{\psi_{1}}_{\ki} + \sqrt{1-\alpha} \ket{\psi_{2}}_{\ki} \right)
\end{equation}
for some $0<\alpha<1$, where $\ket{\psi_{1}}_{\ki}$ and $\ket{\psi_{2}}_{\ki}$ are Gaussian states differing only in the value of $\langle \hat{x} \rangle$. Specifically, $\ket{\psi_{1}}_{\ki}$ and $\ket{\psi_{2}}_{\ki}$ have mean positions $\bar{x}_{0}$ and $\bar{x}_{0}+\Delta x_{0}$ respectively, standard deviation in position $\sigma_{x}$, standard deviation in momentum $\sigma_{p}$, and for simplicity we take both wavepackets to have the same initial mean momentum. This is illustrated in Fig.~(\ref{fSuperposCartoon}). The normalisation factor $N$ is then given by
\begin{equation} \label{eNormFacSuperposition}
N = 1+2 \sqrt{\alpha(1-\alpha)} e^{-\frac{1}{2}\left( \frac{\Delta x_{0}}{2 \sigma_{x}}   \right)^{2}}.
\end{equation}
Note that the exponential factor in Eq.~\eqref{eNormFacSuperposition} is the overlap of the constituent states, $\braket{\psi_{1}\vert\psi_{2}}_{\ki}$ (c.f.~the grey region in Fig.~(\ref{fSuperposCartoon})). 

We denote the average clock time in this case by $\langle \hat{T}_{\cl} \rangle_{\mathrm{sup}}(t)$. To extract the part of the effect arising from quantum coherence, we contrast this with the case of a classical mixture of two such states according to probabilities $\alpha$ and $1-\alpha$. The average clock time in the latter case, which we denote $\langle \hat{T}_{\cl} \rangle_{\mathrm{mix}}(t)$, can easily be calculated via Eq.~(\ref{eQuantTimeDilation}) by taking the corresponding weighted sum, i.e.
\begin{equation}
\langle \hat{T}_{\cl} \rangle_{\mathrm{mix}}(t) = \alpha \langle \hat{T}_{\cl} \rangle_{\psi_{1}}(t) + (1-\alpha) \langle \hat{T}_{\cl} \rangle_{\psi_{2}}(t),
\end{equation}
where $\langle \hat{T}_{\cl} \rangle_{\psi_{i}}(t)$ is the expectation value for state $\ket{\psi_{i}}_{\ki}(t)$.
For simplicity of expression, we consider good clocks in the sense discussed in Sec.~\ref{sQuantClocks}, so that the contribution arising from a nonzero $\hat{E}(t)$ is neglible. One then has
\begin{equation} \label{eCohMixDisc}
\langle \hat{T}_{\cl} \rangle_{\mathrm{sup}}(t) = \langle \hat{T}_{\cl} \rangle_{\mathrm{mix}}(t) + T_\textup{coh}(t),
\end{equation}
where $T_\textup{coh}(t)$, the contribution due to coherence between the two constituent Gaussian states, is given by
\begin{equation} \label{eQuantDiscr}
T_\textup{coh}(t) := \frac{N-1}{N} \left[ \left( \frac{\Delta x_{0}}{2 \sigma_{x}} \right)^{2} \frac{\sigma_{v}^{2}}{c^{2}}-\frac{g \Delta x_{0}}{c^{2}} (1-2\alpha) \right] \frac{t}{2}
\end{equation}
where ${\sigma_{v}:=\sigma_{p}/m}$ is their standard deviation in their initial ``velocity''. We recall that, since the wavepackets saturate the uncertainty relation, we have that $\sigma_{p}=1/(2\, \sigma_{x})$.

In Eq.~(\ref{eQuantDiscr}) we see two contributions to $T_\textup{coh}(t)$: one proportional to $\sigma_{v}^{2}$ and one due to the average difference in gravitational potential experienced by the wavepackets ($g \Delta x_{0}$). In other words, we see separately how motional and gravitational time dilation act on the wavefunction to generate quantum coherence effects. Noting that we can write $g \Delta x_{0} = v_{g}^{2}/2$, where $v_{g}$ is the velocity accrued by a classsical object falling a distance of $\Delta x_{0}$ from rest, we see that for a given $\Delta x_{0}/\sigma_{x}$ and $\alpha$, the relative strength of the motional and gravitational parts is determined by the comparative magnitudes of $\sigma_{v}$ and $v_{g}$. 

Furthermore, considering Eq.~(\ref{eNormFacSuperposition}), one can see that $T_\textup{coh}(t)$ decreases exponentially with increasing $\Delta x_{0}$, a consequence of the decreasing overlap between the two wavepackets. From this we infer that $T_\textup{coh}(t)$ arises due to interference between $\ket{\psi_{1}}_{\ki}$ and $\ket{\psi_{2}}_{\ki}$. On the other hand, we evidently have $T_\textup{coh}(t) = 0$ when $\Delta x_{0} = 0$, as in that case $\ket{\psi_{1}}_{\ki} = \ket{\psi_{2}}_{\ki}$, and we return to the classical-motion scenario described in Sec.~\ref{sClassMotion}. Consequently there exists, for a given $\sigma_{v}$ and $\alpha$, an intermediary value of $\Delta x_{0} / \sigma_{x}$ which maximises $T_\textup{coh}(t)$ by finding a balance between the wavefunction overlap and the ``nonclassicality'' of the kinematic state.

Given that modern-day atomic clocks are accurate and precise enough to observe classical relativistic time dilation in tabletop experiments~\cite{chou2010optical}, it is natural to ask whether the quantum contribution to the time dilation might also be observable. For the purposes of illustration, let us consider an aluminium atom (whose mass is approximately $27$~u). This is inspired by a state-of the-art ``quantum logic clock'' based on optical transitions in a single aluminium ion~\cite{brewer201927}. For clarity, we use the (Van der Waals) radius of aluminium, specifically $r_\mathrm{Al}=184$~pm~\cite{mantina2009consistent}, as a unit of length. Then from Eq.~(\ref{eQuantDiscr}), we see that for a balanced superposition (i.e. $\alpha=1/2$), taking $\sigma_{x}=2$~$r_\mathrm{Al}$ and $\Delta x_{0}=4$~$r_\mathrm{Al}$, we have $T_\textup{coh}(t) \sim 10^{-16}$~s in a laboratory time of $t=1$~s,\footnote{This is well within the coherence times achievable in both trapped-ion and neutral-atom optical atomic clocks~\cite{ludlow2015optical}. Maintaining a coherent superposition of heights for $\sim 1$~s is a separate challenge, achieved in e.g.~\cite{kovachy2015quantum}.} which is well within the measurement capability of state-of-the-art clocks. We have of course, assumed good clocks so that the contribution due to the error term $\hat{E}(t)$, is neglectable. While we have demonstrated that this is a justified approximation for clocks with approximately non-dispersive dynamics, such as the Quasi-Ideal clock, this approximation might not be so well suited to the qubit-phase and Seleker-Wigner-Peres clocks previously discussed. This appears very promising, though we stress that this example is illustrative. A careful analysis concerning potential experimental platforms is required before conclusions can be drawn about the prospect of observing the effect. %We envisage these results will be inspirational for further experimental platform specific, analysis.

\section{Clock Precision and the Coupling of Temporal and Kinematic Degrees of Freedom}
In this section we show how the clock's precision is modified by the relativistic coupling between its kinematic and temporal degrees of freedom. In particular we show how the entanglement generated by the interaction Hamiltonian in Eq.~(\ref{eIntHam}) increases the uncertainty associated with measurements of the temporal variable. 

We quantify the clock's precision via the standard deviation of the clock time, which we denote $\sigmaT{}(t)$. For simplicity, we ignore the contribution due to gravity by setting $g=0$, and consider only temporal measurements which correspond to a projection-valued measure. %, rather than general POVMs. 
In order to calculate the relativistic contribution to $\sigmaT{}(t)$ to leading order, we now include Hamiltonian terms up to order $\hat{p}^4/(m c)^{4}$ (see \App~\ref{app:uncert}). Let $\sigmaT{,\NR}(t)$ denote the standard deviation of the clock time in the absence of the special relativistic coupling. Then, assuming the clock's kinematic and temporal degrees of freedom to be uncorrelated at $t=0$, one finds that $\sigmaT{}(t)$ separates into 
\begin{equation}
\sigmaT{}(t) = \sigmaT{,\NR}(t)+ \sigmaT{,\I}(t) + \sigmaT{,\NI}(t),
\end{equation}
where the term $\sigmaT{,\I}(t)$ is a contribution that remains finite in the case of an idealised clock, given by
\begin{equation} \label{eSigmaTIdeal}
\sigmaT{,\I}(t) := \frac{t^{2}}{8\,\sigmaT{,\NR}(t)} \frac{\langle \hat{p}^{4} \rangle + \sigma_{p^{2}}^{2}}{m^{4}c^{4}},
\end{equation}
where $\sigma_{p^{2}}$ is the standard deviation of the observable $\hat{p}^{2}$, and the term $\sigmaT{,\NI}(t)$ is the relativistic contribution arising due the clock's non-idealised nature. Its full form is given in \App~\ref{app:uncert}. Note that $\sigmaT{,\I}(t) > 0$, and the effect of the relativistic coupling on good clocks (in the sense discussed in Sec.~\ref{sQuantClocks}) is therefore to increase the clock's temporal uncertainty. In the case of an idealised clock, $\sigmaT{,\NR}(t)$ is constant in \exttime{}~time (see \App~\ref{app:POVMs}) and can be made arbitrarily small by an appropriate choice of ${\rho}_{\cl}(0)$. This feature can also be well approximated with the Quasi-Ideal clock. Consequently, according to Eq.~\eqref{eSigmaTIdeal}, the standard deviation of the clock time increases quadratically with the \exttime{}~time in these cases.

\begin{figure}[h]
	\centering
	\includegraphics[width=0.99\columnwidth]{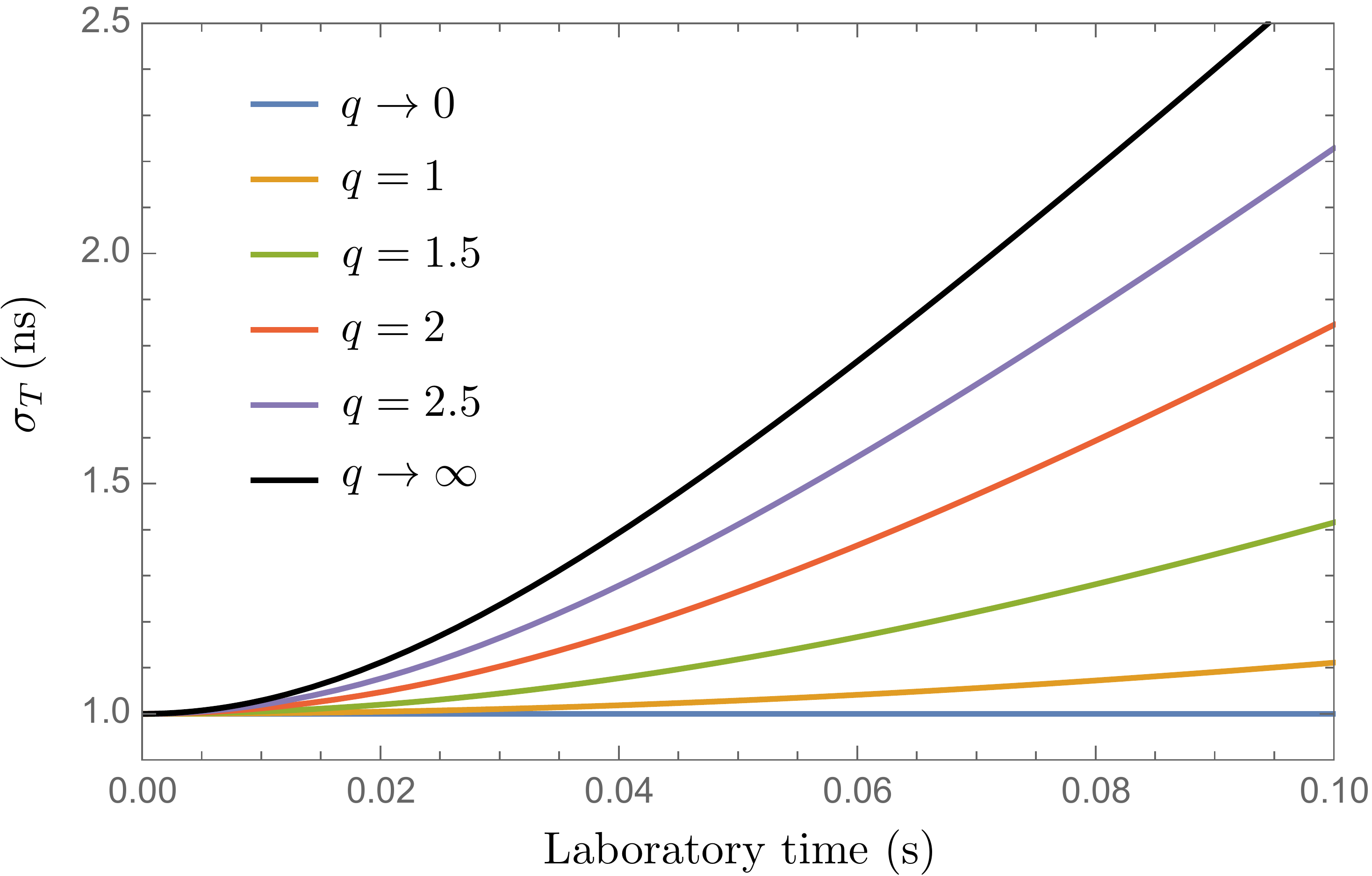}
	\caption{The standard deviation of the clock time, conditioned on discovering that the clock's momentum lies within the central (most likely) bin of variously coarse-grained measurements. The momentum follows a Gaussian distribution centred around $p=0$, and the (idealised) clock is initialised such that its time also follows a Gaussian distribution. For illustration, we have taken $\sigmaT{}(0)=1$~ns, and chosen the other initial parameters to correspond to an electron with $\sigma_{x}=1$~nm. The coarseness of the measurement is determined by $q:=\delta p / \sigma_{p}$, i.e. the ratio of the bin size to the standard deviation in momentum.  } \label{fig:conditioned_x} 
\end{figure}

The decrease in precision is a consequence of the clock's temporal state losing information via its entanglement with the kinematic degrees of freedom. We now show how this effect can be reduced by recovering some of the lost information via a measurement of the clock's momentum. We consider course-grained momentum measurements, and show how the uncertainty in the clock time decreases as the measurement is made more precise. We again choose a classical (i.e. Gaussian) state of motion, as in Sec.~\ref{sClassMotion}, and we consider an idealised clock for simplicity, as this can be approximated arbitrarily well (see the discussion in Sec.~\ref{sQuantClocks}). We define a set of projection operators $\lbrace \hat{\Pi}_{n,\delta p} \rbrace_n$ acting on $\mathcal{H}_{\ki}$, with $n\in \mathds{Z}$ and $\delta p >0$, which correspond to a partition of the range of momentum values into bins of width $\delta p$, with bin~$n$ centred on momentum value $n \, \delta p$, i.e.
\begin{equation}
\hat{\Pi}_{n,\delta p} := \int_{(n-1/2) \delta p}^{(n+1/2) \delta p} d p \ketbra{p}{p}_{\ki}.
\end{equation}
The bin width $\delta p$ therefore characterises the amount of knowledge that we gain from the measurement, and the degree of localisation in momentum space of the post-measurement clock state. As $\delta p$ approaches zero, $\lbrace \hat{\Pi}_{n,\delta p} \rbrace_n$ approaches the set of projectors onto momentum eigenstates. As $\delta p \to \infty$, on the other hand, the projectors tend to the identity operator, i.e. the case where no measurement is performed. Fig. \ref{fig:conditioned_x} shows examples of $\sigmaT{}(t)$ conditioned on a given outcome of the measurement, at different \exttime{}~times $t$, and for different values of $\delta p$. Quantifying the coarseness of the measurement by $q:=\delta p / \sigma_{p}$, we find that for $q\to 0$, we recover all of the information leaked into the kinematic state, i.e. after the measurement, $\sigmaT{}(t)=\sigmaT{}(0)= \sigmaT{,\NR}(t)$. For $q\to \infty$, on the other hand, no information is recovered, as no measurement had been performed.
  
%%%%%%%%%%%%%%%%%%%%%%%%%%%%%%%%%%%%%%%%%%%%%%%%%%%%%%%%%%%%%%%%%
\newpage
\section{Discussion}\label{Discussion}

Throughout this work, we have made reference to the \exttime{} time $t$, which can be interpreted as the time marked by a classical clock in the laboratory frame. One may however wish to consider a fully quantum set-up, without reference to an extrinsic classical time, as in the Page and Wootters formalism~\cite{page1983evolution}, which can be employed in relativistic contexts~\cite{smith2019relativistic,hoehn2020equivalence}. In our framework, one may simply treat $t$ as a bookkeeping coordinate, to be used to calculate the average clock times of multiple quantum clocks, and thus their average clock times relative to each other. In doing so, all our derived results can be written in terms of the average readings of quantum clocks without any reference to a classical clock or extrinsic time.

The effect that we describe in Sec.~\ref{sQuantMotion} is distinct from the manifestation of relativistic time dilation between paths in quantum interferometry experiments. Examples of the latter include phase shifts observed in a neutron matter-wave interferometer~\cite{colella1975observation}, the prediction that redshift-induced which-path information leads to a reduction in interferometric visibility~\cite{zych2011quantum}, and the proposal to observe gravitational phase shifts in an atom interferometer~\cite{roura2020gravitational}. We note that the latter proposal is distinct from the reinterpretation of measurements of atom-interferometry-based accelerometers as a gravitational redshift effect (see~\cite{Peters1999,muller2010precision} and subsequent criticisms~\cite{wolf2010atom,wolf2011does,sinha2011atom}). These works concern the phase accrued between distinct paths in space as a consequence of relativistic time dilation, which is a distinct phenomenon from the one that we consider in this article. Here we ask what time is measured by a clock delocalised across distinct spacetime trajectories; a phase between two spatial components of that delocalisation affects the answer to this question, but it is not itself the quantity of interest.

The present work has been concerned with the effect of relativity in a stopwatch scenario, that is, we are interested in the time elapsed between two events. This contrasts with the ticking clocks discussed in~\cite{erker2014quantum,rankovic2015quantum,erker2017autonomous,woods2018quantum,woodsTC,schwarzhans2020autonomous}. The incorporation of relativity into the latter is an open problem. 

The reduction in clock precision that we predict is both a quantum and relativistic effect, as the general theory of relativity allows for clocks which are always perfectly precise, and clocks in non-relativistic quantum theory are generally uncoupled from their kinematic degrees of freedom. This leads to the interesting statement that relativity requires us to ask how a clock is moving as well as the time it measures, or we must pay a penalty in its precision. Though our explicit calculation of the decreased clock precision was carried out neglecting the effect of gravity, the same principle will hold for a nonzero gravitational field. A general initial clock state which is uncorrelated between temporal and kinematic degrees of freedom will only typically remain in the set of separable states if either the eigenstates of the total Hamiltonian are separable (which they are not), or if the initial state is particularly mixed~\cite{zyczkowski1998volume}. One can therefore only escape the relativistic decrease in precision by increasing the mixedness of the initial state (and thus decreasing its precision anyway). The process of recovering the precision via measurements of the kinematic state will likely be modified by gravity, however. In particular, the gravitational coupling between the temporal degree of freedom and the clock's spatial position will mean that a perfectly accurate measurement of the momentum degree of freedom will no longer totally restore the clock's precision.

The decreased clock precision is related to the effective decoherence of quantum systems discussed in~\cite{pikovski2015universal}, though, contrary to our work, it is the decoherence of the kinematic degrees of freedom which is studied in that case. Other work has predicted a limit of the ability of multiple clocks to jointly measure time due to a gravity-mediated interaction between them~\cite{ruiz2017entanglement}, or focused on special relativistic effects \cite{paige2018quantum}. Furthermore, our work shows that the qubit phase shift seen in~\cite{anastopoulos2018equivalence} is in fact a specific instance of a universal time-dilation effect. We model the effect of mass-energy equivalence in the same manner as these authors, and in particular the resulting relativistic coupling between a system's internal and centre-of-mass degrees of freedom described by the interaction Hamiltonian in Eq.~\eqref{eIntHam} in the limit of low velocities and a weak gravitational field. Outside of this regime, a fully relativistic theory is necessary. An analysis and rebuttal of some of the criticisms of the model can be found in~\cite{Pikovski2015QandA}, including a discussion of the validity of describing relativistic effects via the Schr\"odinger equation. Since a full theory of quantum gravity is yet to be experimentally established, all results at the interface between quantum theory and gravity are naturally controversial. Ultimately, future experiments will determine the full scope of validity for these results.

%%%%%%%%%%%%%%%%%%%%%%%%%%%%%%%%%%%%%%%%%%%%%%%%%%%%%%%%%%%%%%%%%

\section{Conclusion}

In the non-relativistic limit, we characterised a generic quantum clock by an initial state ${\rho}_\cl (0)$, a Hamiltonian $\hat H_\cl$ and a POVM $\{\hat F(s)\}_{s\in\mathcal{S}}$ corresponding to the measurement of the clock time. Not all such systems can serve as clocks; for example, a clock where ${\rho}_\cl (0)$ is an energy eigenstate will not evolve in time. An unavoidable consequence of quantum theory is that any finite-energy clock with perfectly distinguishable temporal states cannot be perfectly accurate (a constraint which is distinct from the uncertainty principle). We quantified the inaccuracy of a generic clock, with arbitrary temporal states, by defining an error operator. We call a clock good, when its error operator is small in norm.

Setting this framework into a relativistic context, we considered the average time dilation according to a generic quantum clock, finding that for Gaussian states of motion, any quantum clock will experience the usual classical relativistic time dilation with some quantum error (given by the trace of the error operator) which can be made arbitrarily small. Taking a nonclassical state of motion, namely a superposition of Gaussian states, we found that interference results in an average clock time which differs from the prediction of classical relativity, even when the error operator of a clock is zero (an idealised clock). Considering the example of an aluminium ion in such a state of motion, we found that this quantum relativistic time dilation discrepancy should be of an observable magnitude.

We then discussed how relativity leads to entanglement between the clock's time-measuring and motional degrees of freedom. We calculated the corresponding reduction in clock precision for classical states of motion, showing how it increases over time. We demonstrated how the information lost from the clock's time-measuring degrees of freedom can be recovered by performing a measurement of the clock's motion in addition to the clock time.
%\vspace{-0.04cm}

It is our hope that these results will fuel interest from the experimental community, so that experiments might be devised to observe combined quantum relativistic effects, perhaps shedding light on the obscure relationship between the two theories.%\\${}$\\${}$\\${}$\\${}$\\${}$%\vspace{1cm}

%%%%%%%%%%%%%%%%%%%%%%%%%%%%%%%%%%%%%%%%%%%%%%%%%%%%%%%%%%%%%%%%%

\section*{Acknowledgments}  
The authors would like to thank Alexander Smith for useful discussions and Yanglin Hu for correcting typos. S.K. acknowledges financial support from ETH Z\"urich during his Master's degree (Birkigt Scholarship).   M.P.E.L. acknowledges support from the ESQ Discovery Grant \emph{Emergent time - operationalism, quantum clocks and thermodynamics} of the Austrian Academy of Sciences (\"{O}AW), as well as from the Austrian Science Fund (FWF) through the START project Y879-N27. M.P.W. acknowledges funding from the Swiss National Science Foundation (AMBIZIONE Fellowship PZ00P2\_179914), the NCCR QSIT, and  FQXi grant \emph{Finite dimensional Quantum Observers} (No. FQXi-RFP-1623) for the programme \emph{Physics of the Observer}.

%%%%%%%%%%%%%%%%%%%%%%%%%%%%%%%%%%%%%%%%%%%%%%%%%%%%%%%%%%%%%%%%%

\bibliographystyle{naturemag}
\bibliography{reltimedil_BiBTeX}

%%%%%%%%%%%%%%%%%%%%%%%%%%%%%%%%%%%%%%%%%%%%%%%%%%%%%%%%%%%%%%%%%

\input{AppendixV4}

\end{document}

%% file: MischaCommands.tex
\usepackage{dsfont}
\usepackage{bbm} % WARNING: JNP has problems with this package when submitting PDFs generated using it. i.e. using the \id command
\usepackage[normalem]{ulem} % For using the \sout{Hellow word} command to strike-out text.
\usepackage{soul} %other strick throuh package use command \st{hello}
\usepackage{color} % for \color commands
\usepackage{braket} %for the \braket and \ketbra commands
\usepackage{amsthm} %for proof environment and more

\usepackage{mathtools} %for Shishir's command
\usepackage[toc,page]{appendix}

\newcommand{\ketbra}[2]{\ket{#1}\!\!\bra{#2}}
\newcommand{\be}{\begin{equation}}
\newcommand{\ee}{\end{equation}}
\newcommand{\nn}{{\mathbbm{N}}}
\newcommand{\rr}{{\mathbbm{R}}}

\newcommand{\zz}{{\mathbbm{Z}}}
\newcommand{\me}{\mathrm{e}}
\newcommand{\mi}{\mathrm{i}}
\newcommand{\id}{\mathbbm{1}} %{\mathds{1}} to be used with \usepackage{dsfont}. Alternativerly use {\mathbbm{1}}  

\newcommand*{\Scale}[2][4]{\scalebox{#1}{\ensuremath{#2}}}

\newcommand{\exttime}{laboratory} % extrinsic/lab/Schroedinger time
\newcommand{\ki}{\textup{\scriptsize{k}}} % kinematic degree of freedom
\newcommand{\cl}{\textup{c}} % Clock space label
\newcommand{\NR}{\textup{\tiny{NR}}}  %{\mathrm{NR}} % Non Relativistic subscript label

\newcommand{\T}{\tiny{T}}

\newcommand{\sigmaT}[1]{\Scale[1.3]{\sigma}_{\T #1}}

\newcommand{\I}{\textup{\tiny{I}}}

\newcommand{\NI}{\textup{\tiny{NI}}}

\def\ba#1\ea{\begin{align}#1\end{align}} % use "\ba" and "\ea to begin/end equation environment "align"

\newcommand{\doc}{\text{article}}  % name of type of document (letter, manuscript, communication etc) Use "\doc~" with no spaces afterwards (if sapce needed, othothwerise used "\doc")
\newcommand{\App}{\text{Appendix}}

\newcommand\mpwS[1]{MW:{\let\helpcmd\sout\parhelp#1\par\relax\relax} }
\long\def\parhelp#1\par#2\relax{%
	\helpcmd{#1}\ifx\relax#2\else\par\parhelp#2\relax\fi%
}

%% file: AppendixV4.tex
\onecolumngrid

\begin{appendix}

\section*{Appendices}
\section{Relativistic coupling of the clock's temporal and kinematic degrees of freedom}\label{app:coupling}

For completeness, we give a derivation of the coupling between the clock's kinematic and time-measuring degrees of freedom, based on the one in~\cite{zych2017quantum}. We use the post-Newtonian metric to approximate a weak gravitational field. In coordinates $\tilde{x}^{\mu}=\left(\tilde{t},\vec{r}\right)$ corresponding to an observer at rest at $r=\infty$, this metric is given by
\begin{equation} \label{eMetricDist}
\begin{aligned}
\tilde{g}_{00}= - \left[ 1 + \frac{2 \Phi(r)}{c^{2}} + \frac{2 \Phi(r)^{2}}{c^{4}} \right] + \mathcal{O} \left( \left( \frac{\Phi(r)}{c^{2}} \right)^3 \right) \quad  \text{ and } \quad \tilde{g}_{ij}= \delta_{ij} \left[ 1 - \frac{2 \Phi(r)}{c^{2}} \right] + \mathcal{O} \left( \left( \frac{\Phi(r)}{c^{2}} \right)^3 \right)  ,
\end{aligned}
\end{equation}
where $r=|\vec{r}|$, $i$ and $j$ label spatial coordinates, $\Phi(r)=-G M/r$ is the gravitational potential of the Earth and $M$ is its mass. Let us now consider the laboratory frame $x^{\mu}=(t,\vec{r})$, where $t$ is the proper time of an observer at rest at $\vec{r}=\vec{r}_{0}$. Transforming the metric in Eq. \eqref{eMetricDist} to the laboratory frame, we have,
\begin{equation} \label{eMetricLab}
g_{00} = \left[ 1 - \frac{\Phi_{0}}{c^{2}} \right]^2 \tilde{g}_{00} + \mathcal{O} \left( \left( \frac{\Phi(r)}{c^{2}} \right)^3 \right) \qquad  \text{ and } \qquad  g_{ij} = \tilde{g}_{ij},
\end{equation}
where $\Phi_{0}:=\Phi(r_{0})$. In the laboratory frame the clock has energy $E_{\mathrm{lab}}$ and spatial momentum $\vec{p}$, with $p=|\vec{p}|$. Now consider the rest frame of the clock, in which it has energy $E_{\mathrm{rest}}$ and zero spatial momentum. The norm of the clock's four-momentum is a scalar quantity, and therefore the same in both frames. Equating this norm in both frames and rearranging, one finds
\begin{equation} \label{eScalarEnergy}
E_{\mathrm{lab}}^{2} = \frac{\bar{g}^{00} E_{\mathrm{rest}}^{2} - p_{j}p^{j} c^{2}}{g^{00}} = -g_{00} \left(E_{\mathrm{rest}}^{2} + p_{j}p^{j} c^{2} \right),
\end{equation}
where $\bar{g}^{00}=-1$ is the $00$-component of the metric in the clock's rest frame, $p_{j}p^{j}=g^{ij}p_{i}p_{j}$ (using the Einstein summation convention), and we have used the fact that $g_{00}=1/g^{00}$. Now, in order to mark the passage of time the clock must have some evolving internal structure, which is associated with some internal energy. Denoting the clock's rest mass by $m$ and its internal energy by $E_{\cl}$, then by mass-energy equivalence,
\begin{equation} \label{eMassEnergyEquiv}
E_{\mathrm{rest}} = m c^{2} + E_{\cl}.
\end{equation}
Let us now restrict to one spatial dimension, namely the radial one, define the coordinate $x:=r-r_{0}$ and take the common approximation of the Newtonian potential around the point $x=0$, i.e. $\Phi(x) \approx \Phi_{0}+g x$, where $g$ is the gravitational acceleration at $x=0$. Then, inserting Eqs.~(\ref{eMetricDist}),~(\ref{eMetricLab}) and~(\ref{eMassEnergyEquiv}) into Eq.~(\ref{eScalarEnergy}), one finds that 
\begin{equation} \label{eScalarEnergy2}
E_{\mathrm{lab}} = mc^{2} + m g x + \frac{p^{2}}{2m} - \frac{p^{4}}{8m^{3}c^{2}} + E_{\cl}  \left( 1 - \frac{p^2}{2m^2c^2}  + \frac{g x}{c^2} \right) ,
\end{equation}
where we have neglected terms of order $(E_{\cl}/mc^{2})^{2}$, $(p/mc)^{4}$, $(g x)^{2}/c^{4}$ and $g x p^{2}/m^{2}c^{4}$ and higher. In the following, we refer to such terms with the notation $\mathcal{O}( 1/c^{4})$. Quantising Eq.~(\ref{eScalarEnergy2}), one finds that the clock evolves subject to the Hamiltonian
\begin{equation} \label{eTotHam}
\hat{H} = \hat{H}_\cl + \hat{H}_\ki + \hat{H}_{\cl\ki},
\end{equation}
where upon quantisation $E_{\mathrm{lab}} \to \hat{H}$ (which acts upon the space $\mathcal{H}_{\cl} \otimes \mathcal{H}_{\ki}$) and $E_{\cl} \to \hat{H}_{\cl}$ (which acts upon the space $\mathcal{H}_{\cl} $),
\begin{equation}
\hat{H}_\ki = mc^{2} + m g \hat{x} + \frac{\hat{p}^{2}}{2m} - \frac{\hat{p}^{4}}{8m^{3}c^{2}}
\end{equation}
represents the rest-mass-energy, gravitational potential energy and kinetic energy (with a lowest-order relativistic correction) of the clock, and
\begin{equation} \label{eAppHck}
\hat{H}_{\cl\ki} = \hat H_\cl\otimes \left( - \frac{\hat p^2}{2m^2c^2}  + \frac{g \hat{x}}{c^2} \right) ,
\end{equation}
is a relativistic coupling of the temporal and kinematic degrees of freedom. We stress that $\hat{H}$ generates evolution with respect to the proper time of a classical observer at $x=0$, i.e. the laboratory time $t$.

\section{Time dilation in generic quantum clocks}\label{app:pertsol}

We now calculate the average time dilation experienced by a generic quantum clock evolving under the Hamiltonian given in Eq.~\eqref{eTotHam}. We begin by employing perturbation theory to calculate the evolution of the clock state in the low-velocity, weak-gravity limit described in \App~\ref{app:coupling} (i.e. neglecting $\mathcal O ( 1/c^4 )$ terms). As in the main text, we choose units such that $\hbar=1$.

Defining a \emph{free} Hamiltonian by $\hat{H}_{0}:=\hat{H}_\cl + \hat{H}_\ki$, so the total Hamiltonian can be written $\hat{H}=\hat{H}_{0}+\hat{H}_{{\cl}\ki}$, where $\hat{H}_{{\cl}\ki}$ is given in Eq.~\eqref{eAppHck}, one can find the evolution of the clock's state via the interaction picture. Denoting the evolution of the state in the interaction picture by $\rho^I(t)$, and following standard perturbation theory (see, e.g. \cite{Shankar1994}), we have that it is related to the Schr\"odinger picture by
\begin{equation}
\rho^I(t)= U_0^\dag(t) \rho(t) U_0(t),\quad  \rho(t)= U(t) \rho(0) U(t)^\dag,
\end{equation}
where $U_0(t)= e^{-i t\hat H_0}$ and $U(t)= e^{-i t \hat H}$. Then $\rho^I(t)$ satisfies,
\begin{equation}
\frac{\partial}{\partial t}\rho^I(t)= \left[ \hat H^I_{{\cl}\ki} (t), \rho^I(t) \right],
\end{equation}
where $\hat H_{{\cl}\ki}^I(t):=e^{i\hat H_{0} t }\hat{H}_{{\cl}\ki} e^{-i\hat H_{0} t } $ is the interaction Hamiltonian in the interaction picture.
A solution to this differential equation can be obtained using the Dyson series. Neglecting terms of order $\mathcal O ( 1/c^4 )$, as in Appendix~\ref{app:coupling}, it yields
\begin{equation}\label{}
\rho^I(t) =\rho(0) - i \,\int_0^t dt^\prime [\hat H_{{\cl}\ki}^I(t^\prime), \rho(0)].
\end{equation}
Thus going back into the Schr\"odinger picture,
\begin{equation}\label{eq:er}
\rho(t) = e^{-i\hat H_{0} t }\rho(0)e^{i\hat H_{0} t } - i \, e^{-i\hat H_{0} t } \left\lbrace \int_0^t dt^\prime [\hat H_{{\cl}\ki}^I(t^\prime), \rho(0)] \right\rbrace e^{i\hat H_{0} t } .
\end{equation}
Tracing over the kinematic space $\mathcal{H}_{\ki}$, we find the reduced state corresponding to the temporal degree of freedom $\rho_\cl(t):=\tr_{\ki}[\rho (t)]$, which we write as
\begin{align}\label{eq:qw}
\rho_\cl(t) = \rho_{\cl,\NR}(t) + \rho_\cl^{(1)}(t) ,
\end{align}
where $\rho_{\cl,\NR}(t):= e^{-i \hat H_\cl t }\rho_{\cl}(0) \, e^{i \hat H_\cl t }$ is the non-relativistic evolution of the reduced state, and the lowest-order relativistic correction to it is given by
\begin{align}\label{eq:relcorr}
\rho_\cl^{(1)}(t) = - i \int_0^t dt^{\prime} \tr_{\ki}\big[e^{-i\hat H_{0} t } [\hat H_{{\cl}\ki}^I(t^\prime),\rho(0) ] e^{i\hat H_{0} t } \big] ,
\end{align}
Now, we consider a clock whose temporal and kinematic degrees of freedom are  initially uncorrelated, writing
\begin{align}
\rho(0) = \rho_\cl(0) \otimes \rho_\ki(0),
\end{align}
for some initial reduced states $\rho_\cl(0)$ and $\rho_\ki(0)$. Then, writing $\hat{H}_{{\cl}\ki}=\hat{H}_{\cl} \otimes \hat{V}_{\ki}$, i.e. $\hat{V}_{\ki}:=- \frac{\hat p^2}{2m^2c^2}  + \frac{\Phi(\hat x)}{c^2}$, we calculate the integrand of Eq.~\eqref{eq:relcorr} to find
\begin{equation}
\tr_{\ki}\big[e^{-i\hat H_{0} t } [\hat H_{{\cl}\ki}^I(t^\prime),\rho(0) ] e^{i\hat H_{0} t } \big] = \left[ \hat{H}_{\cl} , \rho_{\cl,\NR}(t) \right] \, \tr \left[ e^{i\hat H_{\ki} t } \hat{V}_{\ki} e^{-i\hat H_{\ki} t } \rho_\ki(0) \right] .
\end{equation}
Using the relation 
\begin{equation}
e^{\hat X}Ye^{-\hat X}=\hat Y+\left[\hat X,\hat Y\right]+{\frac {1}{2!}}[\hat X,[\hat X,\hat Y]]+{\frac {1}{3!}}[\hat X,[\hat X,[\hat X,\hat Y]]]+\cdots\label{eq:xyx expansion}
\end{equation}
for some $\hat X$ and $\hat Y$ (see Prop. 3.35 in \cite{hall2015lie}), one finds, by neglecting terms of order $1/c^4$ and higher, that %Hadamard's lemma of the Baker-Campbell-Hausdorff formula, one finds that
\begin{equation}
e^{i\hat H_{\ki} t } \hat{V}_{\ki} e^{-i\hat H_{\ki} t } = \frac{1}{c^{2}} \left( - \frac{\hat p^2}{2 m^2} + g \hat x  + \frac{2gt}{m}\hat p - g^2t^{2} \right) .
\end{equation}
Performing the integration in Eq.~\eqref{eq:relcorr}, we find
\begin{equation}
\rho_\cl^{(1)}(t) = - i  \left[ \hat{H}_{\cl} , \rho_{\cl,\NR}(t) \right] \, t R(t),
\end{equation}
where $R(t):= \tr \left[ \left( - \frac{\hat p^2}{2 m^2 c^{2}} + \frac{g \hat x}{c^{2}}  + \frac{\hat{p}gt}{mc^{2}} - \frac{g^2t^{2}}{3c^{2}} \right) \rho_{\ki}(0) \right]$. Having obtained the evolution of the clock's temporal state, one can calculate the average clock time, giving
\begin{equation} \label{eq:GenClockTime}
\langle \hat T_\cl\rangle(t) = \langle \hat T_\cl\rangle_{\NR}(t) + t R(t)\left\{ 1+ \tr\big[\hat{E}(t)\big]\right\}
\end{equation}
where $\langle \hat T_\cl\rangle_{\NR} := \tr \big[ \hat{T}_{\cl}\, \rho_{\cl,\NR}(t)  \big]$ is the clock time in the absence of relativistic effects, and where the error operator $\hat{E}(t) := -i [\hat T_\cl ,\hat H_\cl] \rho_{\cl,\NR}(t) - \rho_{\cl,\NR}(t)$ was introduced in Sec.~\ref{sQuantClocks} of the main text.
%\newpage
\section{Idealised clocks, covariant POVMs and higher moments (in the absence of relativistic effects)}\label{app:POVMs}
In the main text, we defined an idealised clock as one whose first-moment operator satisfies the canonical commutation relations $\big[ \hat{T}_{\cl}^{}, \hat{H}_{\cl} \big]=i\,\id_\cl$ (in the Heisenberg form), and used this to motivate the definition of the error operator $\hat E(t)$ in Eq. \eqref{eEdefn}. We now find covariant POVMs $\{\hat F(s)\}_{s\in\mathcal{S}}$ which are absolutely continuous, that either form an idealised clock, or whose deviation from this condition is covered by a particular form of the error operator $\hat E(t)$. In doing so, we show that our choice of the definition of the error operator is well motivated. We will look at both cases in which $\mathcal{S}$ is bounded or unbounded.

We start with the case that $\mathcal{S}$ is bounded, and denote the boundaries by $\mathcal{S}=[s_0,s_1]$. We expand the domain of $\hat F$ to $\rr$ by periodic extension: $\hat F(s+n(s_1-s_0)):=\hat F(s)$, $s\in\mathcal{S}$, $n\in\zz$.
\begin{comment}
 Namely for $s\in\mathcal{S}$, $\hat F(s)= U_\cl(s) \hat F(0) U^\dag_\cl(s)$, where $U_\cl (s):=  e^{-i \hat H_\cl s }$. While for $s\in\rr$, we can always write $s=s'+n(s_1-s_0)$  for some $n\in\zz$ and $s'\in[s_0,s_1)$. One then defines $\hat F(s):= \hat F(s')$, for all $s\in\rr$ and $n\in\zz$.
\end{comment}

To start with, we will only need the covariance property of the POVMs and the assumption that $\mathcal{S}$ is bounded. Writing the evolution of $\hat T_\cl$ in the Heisenberg picture, and using the covariant property $\hat F(s)= U_\cl(s) \hat F(0) U^\dag_\cl(s)$; $U_\cl (s):=  e^{-i \hat H_\cl s }$, we find for all $t\in\rr$:
\begin{equation}
\begin{aligned}
	\hat T_{\cl}^{}(t)&=U_\cl^\dag (t) \hat T_{\cl}^{} U_\cl(t) = \int_\mathcal{S} ds \, s\, \hat F(s-t) = \int_{s_0}^{s_1} ds \, f(s) \hat F(s-t) \\
	&= \int_{s_0-t}^{s_1-t} ds \,f(t+s)  \hat F(s) =\int_{s_0}^{s_1} ds \,f(t+s)  \hat F(s),\label{eq:dev of T in app}
\end{aligned}
\end{equation}
where $f(s)$ is the sawtooth function: $f(s+n(s_1-s_0)):=s$, $s\in\mathcal{S}$, $n\in\zz$. By taking limits $s_0\rightarrow-\infty$ and/or $s_1\rightarrow+\infty$ in Eq. \eqref{eq:dev of T in app} and noting that POVMs form a resolution of the identity, $\int_{s\in\mathcal{S}} \hat F(s)=\id_\cl$; we find
\begin{align}
	\hat T_{\cl}^{}(t)=
	t\, \id_\cl + \hat T_\cl^{}, %\label{eq:def POVM v1}
\end{align}
for the three unbounded cases: 
\begin{itemize}
	\item [1)] $\mathcal{S}=[s_0, \infty)$ and $t\geq 0$.\quad\qquad 2)  $\mathcal{S}=(-\infty, s_1]$ and $t\leq 0$.\quad\qquad 3) $\mathcal{S}=(-\infty, \infty)$ and $t\in \rr$.
\end{itemize}
%where we have used that the POVMs for a resolution of the identity, $\int_{s\in\mathcal{S}} \hat F(s)=\id_\cl$.
On the other hand, the Heisenberg-picture equation of motion is
%\begin{align}\label{eHeisEOM}
$\frac{d}{dt} \hat T_{\cl}^{}(t)=  - i\, U_\cl^\dag(t)\big[ \hat{T}_{\cl}^{}, \hat{H}_{\cl} \big] U_\cl(t).$
%\end{align}
Thus, equating this with the time derivative of $\hat T_{\cl}^{}(t)=
t\, \id_\cl + \hat T_\cl^{},$ %Eq.~\ref{eq:def POVM v1},
 we arrive at %followed by left and right multiplying by $U_\cl(t)$ and $U_\cl^\dag(t)$, we find
\begin{align}
	\left[ \hat{T}_{\cl}^{}, \hat{H}_{\cl} \right]=i\,\id_\cl\label{eq:weyl form POVM}
\end{align}
for cases 1), 2) and 3) above. Eq. \eqref{eq:weyl form POVM} represents the canonical commutation relations of a specialised Heisenberg form called the Weyl form.

We now investigate the case in which $\mathcal{S}$ is bounded. From Eq. \eqref{eq:dev of T in app} we find
\begin{equation}
\begin{aligned}
	\hat T_{\cl}^{}(t)&= \int_{s_0}^{s_1-s_0-t} ds \,(t+s)  \hat F(s)+ \int_{s_1-s_2-t}^{s_1} ds \,\left(t+s-(s_1-s_0)\right)  \hat F(s) \\
	&= t\, \id_\cl + \hat T_\cl^{}+ (s_1-s_0) \int_{s_1-s_0-t}^{s_1} ds \hat F(s).
	\label{eq:dev of T in app 2}\end{aligned}
\end{equation}
Differentiating under the integral sign, we find 
%\begin{align}
	$\frac{d}{dt} \hat T_{\cl}^{}(t)= \id_\cl+ (s_0-s_1) \hat F(s_1-s_0-t)= \id_\cl+ (s_0-s_1) U_\cl^\dag(t)\hat F(0) U_\cl(t).$
%\end{align}
Hence taking into account Heisenberg-picture equation of motion as before, we obtain
\begin{align}
	\left[ \hat{T}_{\cl}^{}, \hat{H}_{\cl} \right]=i\,\id_\cl+(s_0-s_1)\hat F(0).\label{eq:heinseberg form POVM 1}
\end{align}
The family of quantum states for which the commutator $\big[ \hat{T}_{\cl}^{}, \hat{H}_{\cl} \big]$ is well defined, is the set $\mathcal{S}_\textup{Comm}:=\big\{\rho\in\mathcal{D}(\mathcal{H_\cl})\,\big{|}\,\, \hat T_\cl^{}\rho\in\dom(\hat H_\cl) \,\,\& \,\,\hat H_\cl\rho\in\dom(\hat T_\cl^{}) \big\}$, where $\mathcal{D}(\mathcal{H_\cl})$ is the set of density operators on $\mathcal{H}_\cl$ and $\dom$ is the domain of the self-adjoint operator in question. Thus Eq. \eqref{eq:heinseberg form POVM 1} satisfies the canonical commutation relations on the subset of states $\mathcal{C}:=\{\rho\in\mathcal{S}_\textup{comm}\, |  \, \hat F(0) \rho=0 \}\subseteq \mathcal{S}_\textup{comm}$. If $\mathcal{C}$ is a dense subset of $\mathcal{S}_\textup{comm}$, one says that the pair $\hat T_\cl^{}$ and $\hat H_\cl$ satisfy the Heisenberg form of the canonical commutation relation. The special case of $\mathcal{S}$ bounded and POVMs which are projective values measures, was studied in \cite{garrison1970canonically}, where it was shown that $\mathcal{C}$ is indeed a dense subset of $\mathcal{S}_\textup{comm}$. Another example of a Heisenberg form of the canonical commutation relations is a ``particle in a box'' which is also reviewed in \cite{garrison1970canonically}. These (and other) aspects of covariant POVMs are discussed in detail in~\cite{Alx2019}.  Moreover, our definition of an error operator $\hat E(t)$ in Eq. \eqref{eEdefn} readily accommodates the situation in Eq. \eqref{eq:heinseberg form POVM 1}. In particular, we have $\hat E(t)=  - i(s_0-s_1) \hat F(0) \rho(t)$.

\begin{comment}
Let us start by generalising the definition of $\hat T_\cl$ to include higher moments. Specifically define $\hat T_\cl^{(n)}:= \int_\mathcal{S} s^n \hat F(s)$ for $n\in\nn^0$, whose expectation value is the $n$-th moment of the clock time. Note that $\hat T_\cl^{(0)}=\id_\cl$ since all POVMs form a resolution of the identity by definition, while $\hat T_\cl^{(1)}= \hat T_\cl$ by the definition in the main text. While we will also assume that $\hat T_\cl^{(0)}$ and $\hat T_\cl^{(1)}$ are well defined, not all of the higher moments may be in general. Unless stated otherwise, we will not need to assume all the higher moments are well defined. Note that in the case of a projection-valued measure, we have $\hat T_\cl^{(n)}=\hat T_\cl^{\, n}$.
\end{comment}

We now move on to examine higher moments of covariant POVMs. Assuming all moments up to and including the $n$th moment are well defined, we have
\begin{equation}
\begin{aligned}
	\langle \hat T_{\cl}^{(n)} \rangle (t)&:= \tr\left[\hat T_\cl^{(n)} \rho_\cl(t)\right]= \tr\left[ U_\cl^\dag (t) \hat T_\cl^{(n)} U_\cl(t) \rho_\cl(0)\right] \\
	&= \tr \left[\int_\mathcal{S} ds \, f(s)^n\, U_\cl(s-t) \hat F(0) U_\cl^\dag (s-t) \rho_\cl(0)\right] \\
	&= \tr\left[ \int_\mathcal{S} ds' \, f(s'+t)^n\, U_\cl(s') \hat F(0) U_\cl^\dag (s')\rho_\cl(0)\right] = \sum_{k=0}^n \binom{n}{k} t^{n-k} \langle \hat T_{\cl}^{(k)} \rangle (0),\label{eq:power poly}
\end{aligned}
\end{equation}
where $\langle \hat T_{\cl}^{(0)} \rangle (0)=1$ for all initial states. From this Eq., it follows, for example, that the standard deviation of $\hat T_{\cl}$ is time independent and only depends in the initial state $\rho_\cl$.%It follows that, the variance $\langle \hat T_{\cl}^{(2)} \rangle (t)\,-\,\langle \hat T_{\cl}^{(1)} {\rangle}^2 (t)$ is $t$ independent for all initial clock states. Furthermore, if the initial clock state is localised $ \langle \hat T_{\cl}^{(k)} \rangle (0)=c_0^k$ for some constant $c_0$,\footnote{$\text{For example, $\hat T_\cl^{(n)}=\hat T_\cl^{n}=\int_\rr x^n\ketbra{x}{x}$, and  $\rho_\cl(0)=\ketbra{c_0}{c_0}$ with $\braket{c_0|x}=\delta(x-c_0)$.}$} we find that it reminds so for all times, namely $\langle \hat T_{\cl}^{(n)} \rangle (t)=(t+c_0)^n$.

For an idealised clock, a projection-valued measure is an example of a covariant POVM. To see this, observe that one can write in this case $\hat{F}(s)= \ketbra{s}{s}_\cl$, with $\hat T_\cl^{(1)} \ket{s} = s \ket{s}$. From there, one can use the fact that an idealised clock satisfies $U_\cl^\dag (s) \hat T_{\cl}^{(1)} U_\cl (s) = T_{\cl}^{(1)} + s$ to show that $ U_\cl (s') \ket{s} = \ket{s+s'}$, and therefore $U_\cl(s) \hat F(0) U_\cl^\dag (s) = \hat F(s)$, i.e. the covariance property. If, in addition, the possible clock times span the entire real line (as in the case considered by Pauli~\cite{pauli1958allgemeinen}), then the higher moments satisfy Eq.~\eqref{eq:power poly}, and again the variance is therefore time-independent, regardless of the initial clock state. 

\section{Examples of relativistic time dilation in quantum clocks} \label{app:egs}

\subsection{An idealised quantum clock}\label{app:pertideal}
Recall that an idealised clock is one satisfying the canonical commutation relation, $[\hat T_\cl,\hat H_\cl] = i$, and consequently $\hat{E}(t) = 0$. Let $\hat T_\cl = \hat x_\cl/c+a\id_\cl$ and $\hat H_\cl = c \hat p_\cl$, where $\hat x_\cl$ and $ \hat{p}_\cl$ are position and momentum operators and $a\in\rr$ is chosen such that $\langle \hat T_\cl \rangle(0)=0$. We first consider the case of ``classical'' motion, i.e. where the initial kinematic state of the clock is a pure Gaussian wavepacket with mean momentum $\bar p_0$, standard deviation of the momentum $\sigma_p$ and mean position $\bar{x}_0$, i.e. $\rho_{\ki}(0) = \ketbra{\psi}{\psi}_{\ki}$ with
\begin{align}
\ket{\psi}_{\ki} = \int dp \; \psi (p)\ket{p}_{\ki} ; \qquad \psi(p) := \frac{1}{(2 \pi \sigma_{p}^{2})^{1/4}}  e^{-\left(\frac{p-\bar{p}_{0}}{2 \sigma_{p}}\right)^{2}} e^{-i \bar{x}_{0}(p-\bar{p}_{0})} .
\end{align}
We then have that ${R}(t) = - \frac{\bar{p}_{0}^{2}+\sigma_{p}^{2}}{2 m^{2} c^{2}} + \frac{g \bar{x}_{0}}{c^{2}} + \frac{\bar{p}_{0}gt}{m c^{2}} - \frac{1}{3} \left( \frac{g t }{c} \right)^{2}$ and therefore
\begin{equation}\label{eq:cohid}
\langle \hat T_\cl \rangle(t) = \left[ 1 - \frac{\bar{p}_{0}^{2}+\sigma_{p}^{2}}{2 m^{2} c^{2}} + \frac{g \bar{x}_{0}}{c^{2}} + \frac{\bar{p}_{0}gt}{m c^{2}} - \frac{1}{3} \left( \frac{g t }{c} \right)^{2} \right] t .
\end{equation}
This expression is the average proper time of a classical clock (see Eq.~\eqref{ePropTime} in the main text) with position $\bar{x}_{0}$ and a random momentum following a Gaussian distribution with mean $\bar{p}_{0}$ and standard deviation $\sigma_{p}$. To summarise, an idealised clock in a classical (i.e. Gaussian) state of motion experiences the usual classical relativistic time dilation.

Let us now consider a nonclassical initial kinematic state, specifically a superposition of two Gaussian states, sometimes referred to as a \emph{cat} state. We will contrast this with the case of comparable classical probabilistic mixture of the two states. Let
\begin{align}
\ket{\psi_{j}}_\ki := \int dp \; \psi_{j}(p)\ket{p}_{\ki}; \qquad \psi_{j}(p) := \frac{1}{(2 \pi \sigma_{p}^{2})^{1/4}}  e^{-\left(\frac{p-\bar{p}}{2 \sigma_{p}}\right)^{2}} e^{-i \bar{x}_{j}(p-\bar{p})}; \qquad j \in \{1,2\} .
\end{align}
We now take the initial kinematic state to be in the superposition $\ket{\psi}_\ki =\frac{1}{\sqrt{N}} \big( \sqrt{\alpha}\ket{\psi_1}_\ki + e^{i\theta}\sqrt{1-\alpha}\ket{\psi_2}_\ki \big)$ for some $\alpha \in (0,1)$ and $\theta \in \rr$, and find the average clock time, which we denote $\langle \hat T_\cl \rangle_{\textup{sup}}$, using Eq.~\eqref{eq:GenClockTime}. We then repeat this procedure for the initial kinematic state in the probabilistic mixture, $\rho_\ki(0) = \alpha \ketbra{\psi_{1}}{\psi_{1}} + (1-\alpha) \ketbra{\psi_{2}}{\psi_{2}}$, and denote the average clock time in that case by $\langle \hat T_\cl \rangle_{\textup{mix}}(t)$. From the linearity of the trace, we see immediately that
\begin{align}
\langle \hat T_\cl\rangle_{\textup{mix}}(t) = \alpha\langle \hat T_\cl\rangle_{\psi_1}(t) + (1-\alpha)\langle \hat T_\cl\rangle_{\psi_2}(t),
\end{align}
where $\langle \hat T_\cl\rangle_{\psi_j}(t)$ is the average clock time of a clock with initial kinematic state $\ket{\psi_{j}}_{\ki}$. One then finds that
\begin{align}
\langle \hat T_\cl\rangle_{\textup{sup}}(t) = \langle \hat T_\cl\rangle_{\textup{mix}}(t) + T_{\textup{coh}}(t),
\end{align}
where 
\begin{equation}	\label{eQuantDiscrPhase}
    T_\textup{coh}(t)
:= \frac{N-1}{N} \left[ \left( \frac{\Delta x_{0}}{2 \sigma_{x}} \right)^{2} \frac{\sigma_{v}^{2}}{c^{2}}-\frac{g \Delta x_{0}}{c^{2}} (1-2\alpha) - 2 \frac{\sigma_{v}^{2}}{c^{2}} \Delta x_{0} (\overline{p}-mgt) \tan \theta   \right] \frac{t}{2}
\end{equation}
is a contribution to the clock time arising due to interference between the two Gaussian states constituting the coherent superposition (see main text), and where ${\sigma_{v}=\sigma_{p}/m}$ and $\sigma_{x}=1/2 \sigma_{p}$ are respectively the standard deviation of the velocity and position of the Gaussian wavepackets used to construct the initial state, and $\Delta x_0 = \bar x_2 - \bar x_1$. We recall that $\hbar=1$, and note that the normalisation factor $N$ is given by
\begin{equation}
N = 1+2 \sqrt{\alpha(1-\alpha)} e^{-\frac{1}{2}\left( \frac{\Delta x_{0}}{2 \sigma_{x}}   \right)^{2}} \cos \theta .
\end{equation}
Setting $\theta=0$ in Eq.~\eqref{eQuantDiscrPhase}, one arrives at Eq.~\eqref{eQuantDiscr} in the main text.

\subsection{Salecker-Wigner-Peres clock}\label{app:swpapp}

Consider a $d$-dimensional Hilbert space and a clock Hamiltonian with equally-spaced energy levels, i.e. ${\hat{H}_{\cl}=\sum_{j=0}^{d-1} j \omega \ketbra{e_{j}}{e_{j}}}$, for some $\omega>0$. This corresponds, for example, to the case where the clock is a spin-$j$ system, and then $d=2j+1$. One can obtain a so-called \emph{time basis} $\{\ket{\theta_{m}}\}$ by taking the discrete Fourier transform of the energy eigenstates,
\begin{equation}
\ket{\theta_{m}} = \frac{1}{\sqrt{d}} \sum_{j=0}^{d-1} e^{-i 2 \pi  j m / d} \ket{e_{j}}, \quad m=0,1,\ldots,d-1
\end{equation}
which are used to construct a clock-time operator $\hat{T}_{\cl}=\sum_{m=0}^{d-1} m \frac{T_{0}}{d} \ketbra{\theta_{m}}{\theta_{m}}$, with $T_{0}:=\frac{2 \pi}{ \omega }$. When the clock is in the state $\ket{\theta_{m}}$, we have $\langle \hat T_\cl\rangle= m T_{0}/d$. Choosing an initial state $\rho_{\cl}(0)=\ketbra{\theta_{0}}{\theta_{0}}$, then the clock state will at regular time intervals $t_m:=m T_{0}/d$, $m\in\zz$ ``focus'' into one of the time eigenstates, i.e. $\langle \hat T_\cl\rangle_{\NR}\left(t_m \right)= m T_{0}/d \,(m \text{ mod } d)$. At intermediate times $t\neq t_m$, on the other hand, the clock will be in a superposition of its time eigenstates (see e.g.~\cite{woods2018autonomous}, \App~B). This setup is known as the Salecker-Wigner-Peres clock (or alternatively, the Larmor clock)~\cite{salecker1958quantum,peres1980measurement}.

A straightforward calculation gives $\tr[[\hat T_\cl,\hat H_\cl]\ketbra{\theta_{m}}{\theta_{m}}] = 0$ for all $m=0,1,\ldots,d-1$, and therefore from the definition of the error operator $\hat{E}(t)$, we have $\tr [ \hat{E}(t_m) ] = -1$ for all $m\in\zz$ and for all clock dimensions $d\in\nn^+$. Consequently, when the Salecker-Wigner-Peres clock is in a time eigenstate, Eq.~\eqref{eq:GenClockTime} tells us that $\langle \hat T_\cl\rangle(t_m) = \langle \hat T_\cl\rangle_{\NR}(t_m) $ for \emph{any} state of the clock's kinematic degrees of freedom. To reiterate, this clock is a special case where the average relativistic time-dilation is exactly cancelled by the quantum error of the clock at regular time intervals regardless of how large $d$ is and regardless of the motional state. In this regard, we note that the Salecker-Wigner-Peres clock was used in~\cite{paige2018quantum}, where it was concluded that ``quantum clocks do not witness classical time dilation''.

\subsection{The Quasi-Ideal clock}

Consider the same setting as for the Salecker-Wigner-Peres clock described above, namely the clock Hilbert space, Hamiltonian $\hat{H}_{\cl}$ and clock-time operator $\hat{T}_{\cl}$, but now with the initial temporal state $\rho_{\cl}(0)=\ketbra{\Psi(m_{0})}{\Psi(m_{0})} $, with
\begin{equation}
\ket{\Psi(m_{0})} = \sum_{m=0}^{d-1} g_{m_{0}}(m) \ket{\theta_{m}} ,
\end{equation}
where $g_{m_{0}}(m)$ is a complex Gaussian distribution centred on $m_{0}$ and with standard deviation $\bar\sigma_\cl\in(0,d)$. The choice $\bar\sigma_\cl=\sqrt{d}$ corresponds to the case in which the standard deviation in both the time basis $\{\ket{\theta_{m}}\}$ and energy basis time basis $\{\ket{e_j}\}$ is approximately the same. For other choices, one obtains a clock whose behaviour tends to that of the idealised clock when $\bar\sigma_\cl\to \infty$ and $d/\bar\sigma_\cl\to \infty$ as $d\to \infty$. This is then the \emph{Quasi-Ideal} clock, introduced in~\cite{woods2018autonomous}. We now show that the contribution of the quantum error to the relativistic time dilation, i.e. $\tr [ \hat{E}(t)]$ (see Eq.~\eqref{eq:GenClockTime}), tends exponentially to zero with increasing clock dimensionality. In other words, we show that the relativistic time dilation experienced by the idealised clock can be approximated arbitrarily well by a Quasi-Ideal clock of high enough dimension. Physically speaking, this is because the Quasi-Ideal clock can move at constant ``velocity'' in the time basis with minimal dispersion.

From the definition of the error operator $\hat{E}(t)$, we have
\begin{equation}
\tr [ \hat{E}(t)] = -i \tr \big[  [\hat T_\cl ,\hat H_\cl] \rho_{\cl,\NR}(t) \big]   - 1
\end{equation}
Choosing the initial clock state $\ketbra{\Psi(0)}{\Psi(0)}$, the non-relativistic evolution of the state $\rho_{\cl,\NR}(t)$ is given by,
\begin{align}
\rho_{\cl,\NR}(t) = e^{-i\hat H_\cl t}\ketbra{\Psi(0)}{\Psi(0)}e^{i\hat H_\cl t},
\end{align}
Now, Theorem 8.1 in \cite{woods2018autonomous} states that 
\begin{align}
e^{-i\hat H_\cl t}\ket{\Psi(0)} = \ket{\Psi ( t /T_{0} )} + \ket{\varepsilon},
\end{align}
where the specific form of $\ket{\varepsilon}$ is irrelevant for the present purpose, only that as $d \to \infty$ and $d/\bar\sigma_\cl \to \infty$,
\begin{align}
\sqrt{\braket{\varepsilon | \varepsilon}} = \mathcal O\Big(  \text{poly}(d)\,\big(e^{-\frac{\pi}{4}\frac{d^2}{{\bar\sigma_\cl}^2}}+e^{-\frac{\pi}{4}{\bar\sigma_\cl}^2} \big)  \Big),\label{eq:norm dynamics qi clock}
\end{align}
and furthermore Theorem 11.1 in \cite{woods2018autonomous} states that
\begin{align}
 [\hat T_\cl,\hat H_\cl] \ket{\Psi(m)}  = i \ket{\Psi(m)} + \ket{\varepsilon_{comm}} \quad \forall \, m\in\rr
\end{align}
where, again, the specific form of $\ket{\varepsilon_{comm}}$ is irrelevant, only that as $d \to \infty$,
\begin{align}
\sqrt{\braket{\varepsilon_{comm} | \varepsilon_{comm}}} =\mathcal O\Big(  \text{poly}(d)\,\big(e^{-\frac{\pi}{4}\frac{d^2}{{\bar\sigma_\cl}^2}}+e^{-\frac{\pi}{4}{\bar\sigma_\cl}^2} \big)  \Big).
\end{align}
Using these two theorems, one finds that 
\begin{align}
\left|\tr [ \hat{E}(t)]\right| &=\left| -i \tr \bigg[ \ket{\varepsilon_{comm}} \left\langle {\Psi  ( t /T_{0} ) } \right\vert +  [\hat T_\cl ,\hat H_\cl] \Big( \ketbra{\varepsilon}{\Psi ( t /T_{0} )} +\ketbra{\Psi  ( t /T_{0} )}{\varepsilon}  + \ketbra{\varepsilon}{\varepsilon}  \Big) \bigg]\, \right|\\
&\leq \left|  \tr \Big[ \ket{\varepsilon_{comm}} \left\langle {\Psi  ( t /T_{0} ) } \right\vert\Big]\,\right| + \left|  \tr \Big[ [\hat T_\cl ,\hat H_\cl] \Big( \ketbra{\varepsilon}{\Psi ( t /T_{0} )} +\ketbra{\Psi  ( t /T_{0} )}{\varepsilon}  + \ketbra{\varepsilon}{\varepsilon}  \Big) \Big]\, \right|\label{eq:intermedate 1}  
\end{align}
We can now bound the terms on the r.h.s. of Eq. \eqref{eq:intermedate 1}. To start with
\begin{equation}
\begin{aligned}
\big|\tr[\ketbra{\varepsilon_{comm}}{\Psi(t/T_{0})}]\,\big|&=\big| \braket{\varepsilon_{comm}|\Psi(t/T_{0})}\big| \leq \sqrt{\braket{ \varepsilon_{comm}|\varepsilon_{comm}}} \\
&= \mathcal O\Big(  \text{poly}(d)\,\big(e^{-\frac{\pi}{4}\frac{d^2}{{\bar\sigma_\cl}^2}}+e^{-\frac{\pi}{4}{\bar\sigma_\cl}^2} \big)  \Big)\label{eq:1st trace eq}
\end{aligned}
\end{equation}
Using the Cauchy-Schwarz inequality, $\left|\tr[\hat A\hat B]\right|^2\leq \tr[\hat A^\dag \hat A] \tr[B^\dag B]$ for bounded linear operators $\hat A$, $\hat B$; one finds that for two kets $\ket{a}$, $\ket{b}$
\begin{align}
\Big| \tr\left[ [\hat T_\cl, \hat H_\cl] \ketbra{a}{b}\right]\Big|^2 &\leq - \tr\left[[\hat T_\cl, \hat H_\cl]^2\right]\braket{a|a} \braket{b|b}  \\
&\leq 2\Big(\big| \tr\left[(\hat T_\cl \hat H_\cl)^2 \right]\big|+  \big| \tr\left[\hat T_\cl^2 \hat H_\cl^2 \right]\big|\Big) \braket{a|a} \braket{b|b} \\
&\leq 2\Big(\left\|(\hat T_\cl \hat H_\cl)^2 \right\|+ \left\|\hat T_\cl^2 \hat H_\cl^2 \right\|\Big) \braket{a|a} \braket{b|b} \\
&\leq 4\left\|\hat T_\cl\right\|^2 \left\|\hat H_\cl \right\|^2 \braket{a|a} \braket{b|b}\leq 4 (T_0 d)^2 (d^2 \omega)^2  \braket{a|a} \braket{b|b}\\
&\leq 16 \pi^2 d^6 \braket{a|a} \braket{b|b}.\label{eq:CS used here}
\end{align}
Therefore, using Eqs. \eqref{eq:1st trace eq}, \eqref{eq:CS used here}, and \eqref{eq:norm dynamics qi clock}; one finds from Eq. \eqref{eq:intermedate 1}
\begin{equation}
\left|\tr [ \hat{E}(t)]\right|= \mathcal O\Big(  \text{poly}(d)\,\big(e^{-\frac{\pi}{4}\frac{d^2}{{\bar\sigma_\cl}^2}}+e^{-\frac{\pi}{4}{\bar\sigma_\cl}^2} \big)  \Big)
\end{equation}
for all $t\in\rr$. We therefore find that for any initial clock width of the form  $\bar\sigma_\cl=d_\cl^\eta$ for some fixed $\eta>0$,  $\tr [ \hat{E}(t)]$ is upper-bounded by a quantity which decreases exponentially with dimension. Furthermore, note that for $0<\eta<1/2$, the standard deviation of $\hat T_\cl$ for the initial Quasi-Ideal clock state approaches zero as $d\to \infty$. Thus in the limit of large dimension, the relativistic time dilation (Eq.~\eqref{eq:GenClockTime}) is that of the idealised clock, up to errors that decay exponentially in the dimension.

\subsection{The qubit-phase clock}

Another clock model sometimes found in the literature (e.g.~\cite{zych2011quantum,ruiz2017entanglement,anastopoulos2018equivalence}) uses the phase difference between the two states of a qubit system to estimate elapsed time. This can be considered a simplified model of the Ramsey method sometimes used in atomic clocks (see e.g.~\cite{ludlow2015optical} Sec.~IVA for a discussion of clock interrogation schemes), where it is assumed that the laser is perfectly tuned to the reference frequency. The qubit is put into the initial state $\frac{1}{\sqrt{2}}\left(\ket{0}+\ket{1}\right)$, whose non-relativistic evolution according to the Hamiltonian $\hat{H}_\cl = -\frac{\omega}{2} \ketbra{0}{0} + \frac{\omega}{2} \ketbra{1}{1}$ results in the state 
\begin{equation}
\ket{\theta} := \frac{1}{\sqrt{2}}\left(\ket{0}+ e^{-i \theta} \ket{1}\right)
\end{equation}
where $\theta = \omega t$. The covariant (see Sec.~\ref{sQuantClocks} and \App~\ref{app:POVMs}) POVM corresponding to the estimation of this phase is given by $\{\hat F(\theta)\}_{\theta\in [0,2 \pi )}$ with $F(\theta)=\frac{1}{\pi}\ketbra{\theta}{\theta}$. This measurement corresponds to a maximum-likelihood estimation of $\theta$, and therefore the laboratory time (see e.g.~\cite{holevo1982probabilistic}, Sec.~4.4). One can then directly calculate the error operator:
\begin{equation}
\hat{E}(t) = -( \id_\cl + \sigma_{x} )  \rho_\cl(t) .
\end{equation}
Since $\tr [ \hat{E}(t)] = 1+ \cos ( \omega t)$, we see that this is only a good clock in the sense defined in Sec.~\ref{sQuantClocks} around $\theta=\pi$, which corresponds to the time at which the clock state $\ket{\theta}$ is orthogonal to the initial state. This limitation can be overcome by performing the phase estimation on many copies of the qubit system (see e.g.~\cite{giovannetti2011advances}). Now, carrying this analysis into a relativistic context, Eq.~\eqref{eQuantTimeDilation} gives the average time dilation experienced by the qubit-phase clock as
\begin{equation}
\langle \hat{T}_{\cl} \rangle(t) =  \langle \hat{T}_{\cl} \rangle_\NR(t) +  t R(t) \left[ 2 + \cos ( \omega t) \right] ,
\end{equation}
which is of course equal to the average time dilation experienced by the idealised clock when $\theta=\pi$. %\mpw{In the comparison with experiment, do we want to point out that we need small error op?}

\section{The effect of the coupling on a clock's precision}\label{app:uncert}

We now show how a clock's precision, as quantified by the standard deviation $\sigma_{T}:=\sqrt{\langle \hat{T}_{\cl}^{2} \rangle -\langle \hat{T}_{\cl} \rangle^{2}}$, changes over time as a result of the relativistic coupling between temporal and kinematic degrees of freedom. For simplicity, we now consider flat (Minkowski) spacetime. Here we work to a precision one order higher than before, i.e. neglecting only $\mathcal{O}  \left((\hat{p}/m c )^{6} \right)$ terms. This is because, as we shall see, relativistic effects on the precision of the idealised clock vanish below $\mathcal{O} ( 1/c^4 )$. Repeating the procedure outlined in \App~\ref{app:coupling}, one finds that the relativistic coupling between temporal and kinematic degrees of freedom is now
\begin{equation}
\hat{H}_{\cl \ki} = \hat{H}_\cl \otimes \hat{W}_{\ki} + \mathcal{O} \left( \left(\frac{\hat{p}}{m c} \right)^{6} \right)
\end{equation}
with $\hat{W}_\ki := -\hat{p}^2/2m^2c^2 + 3\hat{p}^{4} / 8m^{4}c^{4}$.

Now, since in this case $[ \hat{H}_{\cl \ki} , \hat{H}_\ki ] = [ \hat{H}_{\cl \ki} , \hat{H}_\cl ] = 0$, we can simply calculate the relevant observables without going into the interaction picture. Using Eq. \eqref{eq:xyx expansion}, we have %Hadamard's lemma, we have
\begin{equation}
\langle \hat{T}_{\cl}^{n} \rangle (t) = \tr \left[ \, \sum_{q=0}^{\infty} \frac{1}{q !} [i (\id_{\ki}+\hat{W}_{\ki}) t]^{q} \, [\hat{H}_{\cl} , \hat{T}_{\cl}^{n} ]_{q}  \right]
\end{equation}
where $[ \cdotp , \cdot ]_{q}$ denotes the $q^\mathrm{th}$-order nested commutator, i.e. $[\hat{A} , \hat{B}]_{0}:=\hat{B}$ and $[\hat{A} , \hat{B}]_{q}:=[\hat{A} , [\hat{A} , \hat{B}]_{q-1}]$ for $q>0$. Applying the Binomial theorem to $(\id_{\ki}+\hat{W}_{\ki})^{q}$ and truncating at the appropriate order, one finds, after some algebra
\begin{equation}
\langle \hat{T}_{\cl}^{n} \rangle (t) = \langle \hat{T}_{\cl}^{n} \rangle_{\NR} (t) +  \tr \Big[ \left\lbrace i t [\hat{H}_{\cl} , \hat{T}_{\cl}^{n}] \otimes \hat{W}_{\ki} - t^{2} [\hat{H}_{\cl} , [\hat{H}_{\cl} , \hat{T}_{\cl}^{n}] ] \otimes \hat{W}_{\ki}^{2}  \right\rbrace \rho_{\NR}(t) \Big]
\end{equation}
where $\rho_{\NR}(t) := e^{-i ( \hat{H}_{\cl} + \hat{H}_{\mathrm{k}} ) t }\rho(0)e^{i (\hat{H}_{\cl} + \hat{H}_{\mathrm{k}} ) t } $ is the non-relativistic evolution of the clock's total state and $\langle \hat{T}_{\cl}^{n} \rangle_{\NR} (t) := \tr [ \hat{T}_{\cl}^{n} \rho_{\cl,\NR}(t) ]$, and where it is understood that $\hat{W}_{\ki}^{2}$ contains an $\mathcal{O}  \left((\hat{p}/m c )^{6} \right)$ term to be neglected. Assuming an uncorrelated initial state as in \App~\ref{app:pertsol}, i.e. $\rho(0) = \rho_\cl(0) \otimes \rho_\ki(0)$, and further defining $\langle \hat{W}_{\ki}^{n} \rangle_{0}:=\tr [\hat{W}_{\ki}^{n} \rho_\ki(0)]$, we have
\begin{equation}
\langle \hat{T}_{\cl}^{n} \rangle (t) = \langle \hat{T}_{\cl}^{n} \rangle_{\NR} (t) +  i t \langle \hat{W}_{\ki} \rangle_{0} \, \tr \big[ [\hat{H}_{\cl} , \hat{T}_{\cl}^{n}] \rho_{\cl,\NR}(t) \big] - t^{2} \langle \hat{W}_{\ki}^{2} \rangle_{0} \, \tr \big[  [\hat{H}_{\cl} , [\hat{H}_{\cl} , \hat{T}_{\cl}^{n}] ] \rho_{\cl,\NR}(t) \big]  .
\end{equation}
Then by a straightforward, if lengthy, calculation, one finally finds that $\sigmaT{}(t)$, separates into 
\begin{equation}
\sigmaT{}(t) = \sigmaT{,\NR}(t)+ \sigmaT{,\I}(t) + \sigmaT{,\NI}(t),
\end{equation}
where $\sigmaT{,\NR}(t)$ is the clock's standard deviation in the absence of relativistic effects, $\sigmaT{,\I}(t)$ is a contribution which remains in the case of an idealised clock, given by
\begin{equation}
\sigmaT{,\I}(t) := \frac{t^{2}}{8 \sigmaT{,\NR}(t)} \frac{ \langle \hat{p}^{4} \rangle + \sigma_{p^{2}}^{2} }{m^{4}c^{4}}
\end{equation}
where $\sigma_{p^{2}}$ is the standard deviation of $\hat{p}^{2}$, and $\sigmaT{,\NI}(t) $ is a contribution arising due to the non-idealised nature of the clock, given by
\begin{equation}
\begin{aligned}
\sigmaT{,\NI}(t) := & \frac{\langle \hat{W}_{\ki} \rangle_{0} \, t}{2 \sigmaT{,\NR}(t)}  \left\lbrace \tr [ (\hat{E}(t)+\hat{E}^{\dag}(t)) \hat{T}_{\cl} ] - 2 \langle \hat{T}_{\cl} \rangle_{\NR} (t) \tr [\hat{E}(t)] \right\rbrace \\
& - \frac{\left( \langle \hat{W}_{\ki} \rangle_{0} \, t \right)^{2}}{8 \sigmaT{,\NR}(t)^{3}}  \left\lbrace \tr [ (\hat{E}(t)+\hat{E}^{\dag}(t)) \hat{T}_{\cl} ] - 2 \langle \hat{T}_{\cl} \rangle_{\NR} (t) \tr [\hat{E}(t)] \right\rbrace^{2} \\
& - \frac{\left( \langle \hat{W}_{\ki} \rangle_{0} \, t \right)^{2}}{2 \sigmaT{,\NR}(t)}  \left\lbrace 2 \tr [ \hat{E}(t) ] + \tr [ \hat{E}(t) ]^{2} \right\rbrace  \\
& - \frac{\langle \hat{W}_{\ki}^{2} \rangle_{0} \, t^{2}}{2 \sigmaT{,\NR}(t)} \Big\lbrace 2 \tr [\hat{E}(t)] + i \tr [ (\hat{H}_{\cl} \hat{e} \, \hat{T}_{\cl} - \hat{T}_{\cl} \hat{e} \, \hat{H}_{\cl} ) \rho_{\cl,\NR}(t) + \hat{H}_{\cl} \hat{T}_{\cl} \hat{E}(t) - \hat{E}^{\dag}(t) \hat{T}_{\cl} \hat{H}_{\cl} ] \\
& \qquad \qquad \qquad + 2 i \langle \hat{T}_{\cl} \rangle_{\NR} (t)  \tr [ \hat{H}_{\cl} (\hat{E}(t) - \hat{E}^{\dag}(t) ) ] \Big\rbrace
\end{aligned}
\end{equation}
where $\hat{E}(t)$ was defined above, and $\hat{e}:=i[\hat{H}_{\cl},\hat{T}_{\cl}]-\id_\cl$. Note that $\sigmaT{,\I}(t) = \mathcal{O}  \left((\hat{p}/m c )^{4} \right)$, which necessitated that we work to one order higher of precision than when we calculated the clock time (\App~\ref{app:pertsol}).

\end{appendix}